\documentclass[times,12pt]{elsarticle}
\usepackage{color,amsmath,amssymb,fullpage,times,graphicx}
\usepackage{xr}
\newcommand{\bs}{\boldsymbol}
\newcommand{\eps}{\varepsilon}

\usepackage[nodots]{numcompress}
\biboptions{comma,square}

\journal{\hspace{-2in}\colorbox{white}{\textcolor{white}{XXXXXXXXXXXXXXXXXXXXXXXXXXXXXXXXXXXXXXXXXXXXXXXXXXXXXXXXXXXXXXXXXXXXXXXXXXXXXXXXXXXXXXXXXXXXXXXXXXXXX}}}

\begin{document}
\begin{frontmatter}

\title{A Statistical Social Network Model for Consumption Data in Food Webs\footnote{On 2013-09-05, a revised version entitled ``A Statistical Social Network Model for Consumption Data in Trophic Food Webs'' was accepted for publication in the upcoming Special Issue {\it Statistical Methods for Ecology} in the journal {\it Statistical Methodology}.}}

\author[label1]{Grace S.~Chiu}
\author[label2,label3]{Anton H.~Westveld}
\address[label1]{Senior Research Scientist \\ CSIRO Computational Informatics (CCI), GPO Box 664, Canberra, ACT 2601, Australia. \\ Corresponding author. E-mail: grace.chiu@csiro.au}
\address[label2]{Assistant Professor in Statistics \\ University of Canberra, University Drive, Bruce, ACT 2617, Australia. E-mail: anton.westveld@canberra.edu.au}
\address[label3]{Associate \\ Statistics Laboratory @ the Bio5 Institute and Statistics Graduate Inter-Disciplinary Program, University of Arizona, Tucson, AZ 85721, USA. \\  E-mail: antonwestveld@email.arizona.edu}

\begin{abstract}
We adapt existing statistical modeling techniques for social networks to study consumption data observed in trophic food webs. These data describe the feeding volume (non-negative) among organisms grouped into nodes, called trophic species, that form the food web. Model complexity arises due to the extensive amount of zeros in the data, as each node in the web is predator/prey to only a small number of other trophic species. Many of the zeros are regarded as structural (non-random) in the context of feeding behavior. The presence of basal prey and top predator nodes (those who never consume and those who are never consumed, with probability 1) creates additional complexity to the statistical modeling.  We develop a special statistical social network model to account for such network features. The model is applied to two empirical food webs; focus is on the web for which the population size of seals is of concern to various commercial fisheries.
\end{abstract}


\end{frontmatter}

\newpage
\normalsize
\section{Introduction: food webs as social networks}\label{sec-intro}
A food web is a network of organisms and what they consume. When the relationship among these is of interest, organisms (inanimate or otherwise) are typically aggregated at various resolutions to form {\it trophic species}. For example, one trophic species in a given web may consist of various types of organic dead matter collectively referred to as detritus, while another in the same web may consist of the single anole species {\it A.~gingivinus}. We refer to a trophic species as a food web {\it node}. In the trophic context, feeding relations among nodes are of interest. (In contrast, the ecosystem context of a food web concerns the closed-loop transfer of energy and nutrients \citep{christian,chiu.fw,ena,rooney}.) Thus, trophic food web research concerns the understanding of the interdependency, or network structure, among trophic species with respect to predation or consumption behavior.

For a given pair of nodes $(i,j)$, the four possible within-pair trophic relations are depicted as
\begin{equation}\label{link}
i\ \ \ \ \ j\hspace{1in} i\rightarrow j\hspace{1in}i\leftarrow j\hspace{1in} i\leftrightarrow j
\end{equation}
where, conventionally, any link/edge points from prey to predator. From left to right in (\ref{link}), the depictions respectively represent no predation between $i$ and $j$, predation of $i$ by $j$ but not {\em vice versa}, predation of $j$ by $i$ but not {\em vice versa}, and mutual predation between $i$ and $j$. To represent (\ref{link}) in a quantitative framework, ``$i\ \ \ j$\,'' consists of two zero links, each of ``$i\rightarrow j$\,'' and ``$i\leftarrow j$\,'' consists of a zero and a positive link, and ``$i\leftrightarrow j$\,'' consists of two positive links. Thus, each pair $(i,j)$ yields two directed links: from $i$ to $j$, and from $j$ to $i$. Extending this to all $n$ nodes in the food web, we have a network that consists of 2$\times$($n$-choose-2) = $n(n-1)$ pairwise or {\em dyadic} directed links. Presence and absence of feeding interactions are represented by binary links, while consumption volumes are represented by non-negative {\it weighted} or {\it valued} links. A binary link qualitatively describes the pairwise relation, while a weighted link reflects the degree of one node's dominance over the other. Thus, binary and weighted links yield different insights into the network~structure.

Research on network structure arises in many practical settings, commonly in the social sciences, e.g.,~pair bonds \citep{pavel}, international trade and militarized disputes \citep{ward,westveld}. Many quantitative social network analysis (SNA) techniques had been developed to understand network relational patterns \citep{mucha,wasserman}, and some, adopted in food web research \citep{dambacher,krause}. These earlier SNA methods for food webs were largely based on the mathematical notion of equivalence class for defining congruence among individual elements in a given set according to certain criteria (see, e.g.,~\citep{duentsch}); the objective is to seek optimal partitions of the network into compartments of nodes subject to the given criteria. For example, compartments identified in a food web may correspond to trophic levels. \citet{chiu.net} overviews some common SNA techniques for food web research. 

Statistical approaches for SNA began more than half a century ago \citep{goldenberg}, but research interest and methodological advancement in the area grew dramatically in the past two decades, due in part to the exponential increase in computational technologies and the general public's interest in social networks. In particular, statistical regression methodologies were developed only recently to express network links $y_{ij}$ as the random response of within-node and inter-node characteristics \citep{gill2,hoff,hoff.et.al,vanduijn,westveld}. The resulting conclusions about network features are purely empirical and entirely based on observed network attributes without the use of network dynamics or subject-matter theory. \citet{pnas} demonstrate that, in the context of binary food webs, taking a statistical SNA approach can provide an alternative perspective of trophic relational patterns according to feeding activity and preference. These authors adapted the statistical SNA latent-space modeling framework by \citet{hoff} and \citet{ward} to regress the presence-absence of pairwise predation, in a mixed-effects logistic model, on the dyadic characteristic of phylogenetic similarity; eight food webs were analyzed this way.
The basis of their statistical model for SNA is a two-way analysis-of-(co)variance (ANO(CO)VA) model (see Section \ref{sec-binary}). 

\citet{chiu.net} and \citet{pnas} discuss the various advantages of using this statistical SNA framework to study food webs, most notably: The inherent uncertainty in network links is readily acknowledged and modeled by the framework; in addition to ``Who tends to eat whom?'' the model can address ``Why?'' through covariates, avoiding post-SNA ``detective work'' that is typical of conventional SNA methods; rigorous quantitative inference (including predictive inference) can be made for various features of feeding behavior, from food web connectance to the relative distinction among nodes with respect to consumption activity level and preference; and finally, all these can be achieved through a single unified statistical analysis through the regression framework, thus avoiding the intractable propagation of uncertainty in conventional multi-step analyses. 

\begin{figure}[!h]
\begin{center}
\caption{\label{fig-hist} Distribution of consumption data for the Benguela food web. That for the St.~Martin web is very similar in shape.}
\makebox{\includegraphics[scale=.5]{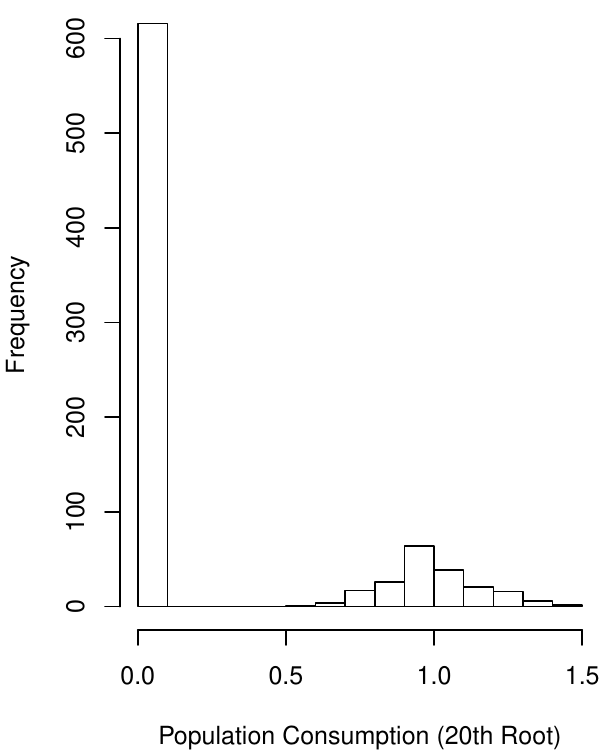}}
\end{center}
\end{figure}

{%
These merits are true for $y_{ij}$ in general, i.e., for binary or weighted data ($y_{ij}$$\ge$0). However, weighted food web data often pose a modeling challenge, due to the high incidence of $y_{ij}$=0. For example, each of the eight food webs analyzed in \citep{pnas} consists of between 69\% and 98\% zeros. %
}%
{%
Direct application of existing statistical SNA techniques to such weighted data would require a special distributional assumption for $y_{ij}$ to account for its extreme point mass at 0 and its continuous distribution away from 0 (Fig.~\ref{fig-hist}). Extending the framework of \citep{pnas} to incorporate a mixture distribution may seem reasonable, but at the expense of model complexity and computational burden. Moreover, almost all observed 0s are biologically deterministic (see Section \ref{sec-magnitude}). This prompted us to propose in this paper an approach that does not require the same level of model complexity as a mixture model, and is reasonably straightforward to implement.%
}

{%
In Section \ref{sec-binary}, we review the latent-space modeling framework used in \citep{pnas} for binary trophic food webs. For consumption data, we propose in Section \ref{sec-magnitude} new components required to handle weighted data with latent-space modeling, and discuss proper interpretation of model parameters from the modified framework. The new framework is demonstrated through two case studies, applied respectively~to the Benguela and St.~Martin food webs. Section \ref{sec-imp} describes the datasets, and presents the Bayesian model hierarchy for each case study. Analysis results are presented and interpreted in Section \ref{sec-results}. We end the paper with a discussion in Section \ref{sec-discuss} on (a) how our proposed approach mitigates some long-standing limitations of popular food web analysis methods, and (b) future improvements to our approach towards handling data from large weighted food webs and their temporal features.%
}
\section{Modeling binary trophic food webs}\label{sec-binary}
{%
Although feeding behavior may be observed at the organism or species level, trophic food web data are typically recorded for $n$ predefined nodes and presented in a square diet matrix $[y_{ij}]$, where $i$ is prey and $j$ is predator, for $i,j=1,\ldots,n$. For binary $y_{ij}$s, the generalized linear mixed ANOCOVA model employed in \cite{pnas} is%
}%
\par\noindent%
\begin{equation}\label{logistic}
y_{ij}|p_{ij}\sim\text{Bernoulli}(p_{ij})\,,\ \ \ \ \ 
\log\frac{p_{ij}}{1-p_{ij}}=\beta_0+\beta_1 x_{ij}+s_i+r_j+\bs{u}_i'\bs{v}_j+\eps_{ij}\,,\ \ \ \ \ \ i\neq j
\end{equation}
{%
where $p_{ij}$ is the probability that $j$ predates on $i$, the optional covariate $x_{ij}=x_{ji}$ is the phylogenetic similarity between $i$ and $j$, $s_i$ is the $i$th random sender (prey) effect, $r_j$ is the $j$th random receiver (predator) effect, $\bs{u}_i'\bs{v}_j$ is the random interaction between $i$ and $j$ expressed as an inner product of $k$-dimensional vectors $\bs{u}_i$ and $\bs{v}_j$, and $\eps_{ij}$ is the remaining random component unattributable to $x_{ij},s_i,r_j,\bs{u}_i$, or $\bs{v}_j$. All of $s_i, r_j, \bs{u}_i, \bs{v}_j,$ and $\eps_{ij}$ are mean-zero Gaussian random errors. Note that expressing the interaction term as the inner product of latent vectors $\bs{u}_i$ and $\bs{v}_j$ is due to \citet{hoff.et.al}. The latent $\bs{u}$- and $\bs{v}$-spaces are abstract entities; their dimension $k$ can be regarded as a model parameter to be estimated from the data, but considered in \citep{pnas} as fixed at $k$=2 for convenient visualization. We discuss the interpretation of the latent spaces in Section \ref{sec-interpret0}.%
}

{%
Complex network dependence not addressed by the ANOCOVA equation (\ref{logistic}) is modeled through%
}%
\begin{equation}\label{cov1}
\mbox{Cov}(s_i,r_i)=\bs{\Sigma}=\left[
\begin{matrix}
\sigma_s^2 & \rho_{sr}\sigma_s\sigma_r \\
\rho_{sr}\sigma_s\sigma_r & \sigma_r^2
\end{matrix}
\right]\hspace{1cm}\mbox{for all }i=1,\ldots,n\,,
\end{equation}
\begin{equation}\label{cov2}
\mbox{Cov}(\eps_{ij},\eps_{ji})=\bs{\Omega}=\sigma^2\left[
\begin{matrix}
1 & \rho \\
\rho & 1
\end{matrix}
\right]\hspace{1in}\mbox{for all }i\neq j\,.
\end{equation}
{%
Eq.~(\ref{cov1}) stipulates that sender and receiver effects due to the same node are possibly correlated ($\rho_{sr}$), with potentially distinct amounts of uncertainty ($\sigma_s$ and $\sigma_r$); this within-node structure is constant across nodes. Eq.~(\ref{cov2}) allows for potential reciprocity between $i$ and $j$ through $\rho$ (constant across all $(i,j)$-pairs), with the typical assumption of homogeneous random errors through $\sigma$. This implies that, conditioned on covariates and sender- and receiver-specific random effects, $y_{ij}$s are still possibly dependent; this dependence is assumed to be solely in the form of reciprocity, i.e., $y_{ij},y_{ji}$ may be conditionally dependent, but $y_{ij},y_{k\ell}$ are conditionally independent for mutually distinct $i,j,k,\ell$ as implied by a Markov graph \citep{frank.strauss,pavel3,pavel,robins}.
}

\subsection{Interpretation of model parameters}\label{sec-interpret0}
{%
Given $i$, the bivariate random effect $[s_i,r_i]'$ describes the level of feeding activity of the node, while adjusting for covariate effects. Feeding activity for $i$ is related to its {\em in-degree} ($\sum_j y_{ji}=$ total activity as predator) and {\em out-degree} ($\sum_i y_{ij}=$ total activity as prey). In- and out-degree unexplained by covariate(s) are captured respectively by $s_i$ and $r_i$. A graph of the estimated $[s_i,r_i]'$ vectors, referred to as Graph SR (see \citep{pnas}), displays the positions of nodes according to their feeding activity; any identifiable cluster may be considered a ``trophic level'' from the perspective of feeding activity as prey and as predator.%
}

{%
Estimates of the $k$-dimensional vectors $\bs{u}_i$ and $\bs{v}_j$ may also be graphed, referred to as Graphs~U and V (see \citep{pnas}); $k$ can be fixed at 1 or 2 
to reduce model complexity and allow easy visualization of the vectors. Alternatively, if $k>$ 2 is deemed appropriate, then the value of $k$ can be determined via optimization criteria and projected onto $\mathbb{R}^2$ for graphical display \citep{hoff,ward}. In either case, the latent $\bs{u}$- and $\bs{v}$-spaces can be regarded, respectively, as preference of being consumed and that of consuming \citep{chiu.net,pnas}. Specifically, if $\bs{u}_i$ and $\bs{u}_j$ are neighbors in the $\bs{u}$-space, then the sending behavior of $i$ to $\ell$ --- while accounting for sending activity --- is similar~to that of $j$ to $\ell$, for all nodes $\ell$. In a food web, this phenomenon roughly translates to $i$ and $j$ being similarly preferred as prey. The same interpretation applies to $\bs{v}_i$ and $\bs{v}_j$ being neighbors in the $\bs{v}$-space, except for their similarity in receiving, or preference for prey. Thus, clustering in the $\bs{u}$- / $\bs{v}$-space suggests trophic levels with respect to preference of being prey/predator. \citet{pnas} demonstrate that trophic clusters identified in Graphs SR, U, and V can differ~substantially depending on the perspective from which trophic relations are viewed. For example,~those nodes that show similarity in feeding activity often differ drastically in their preference for prey.%
}

{%
Finally, the sign of $\rho_{sr}$ corresponds to the trend of Graph SR. A positive trend suggests that (in)active predators tend to be (in)active as prey, and a negative trend suggests that active predators tend to be inactive as prey and {\it vice versa}. Similarly, the sign of $\rho$ corresponds to the tendency of reciprocity of predation, or predator-prey role reversal.%
}

\section{Modeling trophic consumption data}\label{sec-magnitude}
{%
Beyond binary trophic relations, food web consumption data $y_{ij}$ can be, for example, the frequency count (discrete) or the total biomass (continuous) of $i$ being consumed by $j$. Like \cite{pnas}, here we do not model $y_{ii}$. While cannibalism is not rare in nature, we regard it as a given node's self influence, which is typically considered in ecosystem-level food web studies (nutrient or energy flow within a closed system). %
}

{%
Consider the nature of zeros in food web data. It differs from the social context of, say, friendship in which typically (i) non-zero links between two nodes are common, and (ii) the randomness inherent in the links is substantial enough that under different scenarios, zero links can plausibly become non-zero and {\em vice versa}. In contrast, zero links often dominate a food web; given such a link, biological theory can easily identify the nature of this 0, such as the case of a herbivorous grazer almost surely not consuming other animal species under any realistic scenario. Among the vast number of zeros, those regarded as truly random are typically no more than a handful, if any. For this reason, here we regard all observed zeros as {\em structural zeros}, and remove them from consideration when constructing the social network model. Note that for binary food web data, including the zeros in the logistic model as is done in \citep{pnas} implies a different interpretation than what we propose in this paper. Specifically, the complete food web (the set of all $n(n-1)$ directed links) is regarded in \citep{pnas} as one random entity, whereas in our current context, the randomness in each positive directed link is being modeled. By removing all zeros, we not only avoid the complexity required to model the randomness of the complete weighted food web for which $y_{ij}$ has a highly non-standard distribution; additionally, we can substantially reduce computational burden due to a greatly reduced food web size $n$. On the other hand, special care is required, as discussed below, to handle model degeneracies due to undefined sender- and receiver-specific random effects.%
}

\subsection{Special handling of the reduced web}
{%
First, some notation will aid the discussion in this subsection. Define the following sets~of~pairs:%
}%
\par\noindent%
\begin{align*}
{\cal S}^\ast &= \{(i,j):\ j>i,\ \ i,j=1,\ldots,n\} = \text{all $n$-choose-2 pairs in the food web,} \\
{\cal S}_0 &= \{(i,j)\in{\cal S}^\ast: y_{ij},y_{ji}=0\} = \text{all unlinked pairs,} \\
{\cal S} &= {\cal S}^\ast\setminus{\cal S}_0 = \text{all linked pairs,} \\
{\cal S}_1 &= \{(i,j)\in{\cal S}: y_{ij}>0,\ y_{ji}>0\} = \text{all mutually predatory pairs,} \\
{\cal S}_2 &= \{(i,j)\in{\cal S}: y_{ij}>0,\ y_{ji}=0\} = \text{all send-only pairs ($j$ predates on $i$ but not {\em vice versa}),} \\
{\cal S}_3 &= \{(i,j)\in{\cal S}: y_{ij}=0,\ y_{ji}>0\} = \text{all receive-only pairs ($i$ predates on $j$ but not {\em vice versa}).}
\end{align*}
{%
Note that ${\cal S}_0,\ldots,{\cal S}_3$ are disjoint.%
}

{%
We refer to the weighted food web excluding zero links as the ``reduced weighted food web,'' or ``reduced web'' for short. By definition of the reduced web, the set of unlinked pairs ${\cal S}_0$ is discarded. Thus, $\cup_{k=1}^3{\cal S}_k={\cal S}$.%
}

{%
Next, define the following sets of node labels:%
}%
\par\noindent%
\begin{align*}
{\cal I} &= \{1,2,\ldots,n\} = \text{all nodes of the complete food web,} \\
{\cal I}_1 &= \left\{i\in{\cal I}: \sum_{j=1,\ j\neq i}^n y_{ij}>0,\ \sum_{j=1,\ j\neq i}^n y_{ji}=0\right\} = \text{all basal nodes (only consumed but never predate),} \\
{\cal I}_3 &= \left\{i\in{\cal I}: \sum_{j=1,\ j\neq i}^n y_{ij}=0,\ \sum_{j=1,\ j\neq i}^n y_{ji}>0\right\} = \text{all top predators (only predate but never consumed),} \\
{\cal I}_2 &= \left\{i\in{\cal I}: \sum_{j=1,\ j\neq i}^n y_{ij}>0,\ \sum_{j=1,\ j\neq i}^n y_{ji}>0\right\} = \text{all ``middle'' nodes (neither basal nor top predators).}
\end{align*}
{%
Thus, ${\cal I}$=$\cup_{k=1}^3{\cal I}_k$ consists of all $n$ nodes of the food web, partitioned into disjoint sets ${\cal I}_1,{\cal I}_2,$ and~${\cal I}_3$. %
}

{%
For a given node, the standard latent space model employed in \citep{pnas} simultaneously considers links into and out of the node through (\ref{cov1}) and (\ref{cov2}). Yet, for the reduced web, (\ref{cov1}) only applies to ${\cal I}_2$.
In contrast, nodes in ${\cal I}_1$ and ${\cal I}_3$ are linked to other nodes in one direction only, so that (\ref{cov1}) is degenerate and reduces to (a) $s_i$ being independent and identically distributed (iid) as N(0, $\sigma_s^2$) for all $i\in{\cal I}_1$, and (b) $r_i$ being iid N(0, $\sigma_r^2$) for all $i\in{\cal I}_3$. Similarly, (\ref{cov2}) only applies to $(i,j)\in{\cal S}_1$. For ${\cal S}_2$ or ${\cal S}_3$, one of the two links is missing, so that (\ref{cov2}) is degenerate and reduces to $\eps_{ij}$ being~iid~N(0,~$\sigma^2$).%
}

{%
Under this reduced framework, technically $s_i$ is undefined for $i\in{\cal I}_3$ and $r_i$, for $i\in{\cal I}_1$. To facilitate the visualization of feeding activity, we arbitrarily define%
}%
\begin{equation}\label{undefined}
s_i\equiv -4\sigma_s^2\ \ \ \ \ \mbox{for all }i\in{\cal I}_3\,,\hspace{1in}
r_i\equiv -4\sigma_r^2\ \ \ \ \ \mbox{for all }i\in{\cal I}_1
\end{equation}
{%
so that given the variability of overall feeding activity, the ``random effects'' in (\ref{undefined}) are in fact constant, and are appropriately located in the far left tail of the distribution of $s_i$ and $r_i$. Then, any $[s_i,r_i]'$ for $i\in{\cal I}$ may be displayed on the $sr$-plane.%
}

{%
On the other hand, leaving $\bs{u}_i,\bs{v}_i$ for $i\notin{\cal I}_2$ undefined would not hinder the visualization of feeding preference, since we consider the $\bs{u}$- and $\bs{v}$-spaces separately (as opposed to the cross product of the $s$- and $r$-spaces). Thus, we can consider the distribution of $\bs{u}_i$ for all $i\in{\cal I}_1\cup{\cal I}_2$ in the $\bs{u}$-space, and that of $\bs{v}_i$ for all $i\in{\cal I}_2\cup{\cal I}_3$ in the $\bs{v}$-space. As \citet{pnas} do, we also take $k=$ 2 to reduce model complexity, and take Var($u_{iq}$) $=\sigma_{uq}^2,$ and Var($v_{iq}$) $=\sigma_{vq}^2$ for all $i$ and $q=$ 1, 2, where $\bs{u}_i=[u_{i1},u_{i2}]'$ and $\bs{v}_i=[v_{i1},v_{i2}]'$. In Section \ref{sec-imp}, we will consider if $k<$ 2 may be preferred for our case studies.%
}

\subsection{Interpreting model parameters for the reduced web}\label{sec-interpret}
{%
Recall from Section \ref{sec-interpret0} the interpretation of parameters in the model for a complete food web. Here, removing all $y_{ij}$=0 from consideration lead to model degeneracies. Thus, quantities in the reduced framework should be interpreted with extra care. Specifically, the comparison of feeding activity among nodes is made relative to true activity only, so that non-activity is not part of the comparison. Indeed, as shown in the following case studies with consumption magnitude data for reduced webs, Graph SR (e.g.,~Fig.~\ref{fig-sr-adult}) and the inference for $\rho_{sr}$ can suggest that active consumers (those who consume large volumes) are themselves more actively consumed by volume. The inference for $\rho$ can also suggest a high tendency of predator-prey role reversal. These positive correlations may appear counterintuitive for a complete food web which regards non-activity as a description of activity in general. Yet, when non-activity is ignored in the case of a reduced web, ``predator-prey role reversal'' refers to the reciprocation of similar, strictly positive volumes of predation.%
}

{%
Note that whether it be a complete or reduced web, the above interpretation of random effects and correlation parameters corresponds to covariate effects being accounted for.%
}

\section{Modeling the Benguela and St.~Martin reduced webs}\label{sec-imp}
{%
We focus on the well-studied Benguela web, a marine ecosystem off the southwestern African coast, consisting of 29 nodes (Table~\ref{nodes}). It was originally studied by \citet{yodzis} and revisited by others (e.g.,~\cite{pnas,dunne}). Another well-studied web is the terrestrial St.~Martin ecosystem in the Caribbean, with 44 nodes (Table~\ref{nodes2}). We consider the St.~Martin web mainly to demonstrate the behavior of our methodology rather than from subject-matter interest. In contrast, we attempt to address various scientific concerns about the Benguela ecosystem through our methodology.%
}
\subsection{Benguela data}\label{sec-data}
{%
The diet matrix in \citep{yodzis} consists of diet percentages, and is accompanied by data for other variables: adult individual body mass, annual harvest, carrying capacity, ingestion factor, and population biomass. (\citet{yodzis} also compute and provide population biomass adjustments, which we do not consider.) The $(i,j)$th cell of the diet matrix represents the percentage of $j$'s diet through consuming $i$. Thus, each column of the diet matrix necessarily sums to 100 if all organisms in the actual food web are represented in the diet matrix. However, this appears not to be the case, as column sums range from 90 to 131.5 for non-basal nodes.%
} 

{%
To derive weighted data that corresponds to consumption volume, first we scaled the diet percentages according to the column sums so that each scaled column for a non-basal node sums to 100. The columns for the two basal nodes necessarily consist of all 0s and remain unaltered. Next, we used the scaled diet proportions to define two different measures of consumption volume:%
}%
\par\noindent%
\begin{align*}
\begin{split}
(i,j)\text{th population consumption} &=
    (j\text{'s population biomass})\times (j\text{'s ingestion factor})\times \\
 & \hspace{1in} (j\text{'s diet proportion due to }i)
\end{split}
\\
\begin{split}
(i,j)\text{th per-adult consumption} &=
    (j\text{'s adult individual biomass}) \times (j\text{'s ingestion factor})\times \\
 & \hspace{1in} (j\text{'s diet proportion due to }i)
\end{split}
\end{align*}
where the ingestion factor $j$ is its ``fraction of physiologically maximal ingestion'' \citep{yodzis}. Both definitions refer to the biomass of $i$ consumed by $j$. We take%
\par\noindent%
\begin{align}
y_{ij} &=\left((i,j)\text{th population consumption}\right)^\frac{1}{20} \label{20th} \\
\text{or\hspace{5mm}} y_{ij} &= \left((i,j)\text{th per-adult consumption}\right)^\frac{1}{10}\,. \label{10th}
\end{align}
{%
As can be seen in Fig.~\ref{fig-hist}, the 20th-root transformation in (\ref{20th}) results in a reasonably Gaussian distribution after the removal of 0 links. The same is true for the 10th-root transformation in (\ref{10th}) (not shown). The objective is to express either definition of $y_{ij}$ as the response of covariates in the reduced SNA framework. Although these transformations may appear to be drastic and not easily interpretable, having approximate normality eliminates extra model complexity.%
}

{%
Of the variables with data that accompany the matrix of diet percentages in \cite{yodzis}, only four are suitable as covariates, namely, $t_{i1}^s$ = sender adult individual biomass, $t_{i2}^s$ = annual harvest of sender, $t_{j1}^r$ = receiver adult individual biomass, and $t_{j2}^r$ = annual harvest of receiver. The remaining variable, carrying capacity, is not a suitable covariate, as it is highly related to the notion of ingestion factor, which is used to define the response variable. Note that $t^s$s are sender-specific covariates, and $t^r$s are receiver-specific. In addition, we create a pair-specific covariate, $t_{ij}^p$ = taxonomic distance, by comparing the taxonomic classification of $i$ and $j$ according to their domain, kingdom, phylum, class, order, family, genus, and species. This is the distance analog of the conservative phylogenetic similarity measure $z_{ij}$ of \cite{pnas} prior to log-transformation. Altogether, we have five covariates, all of which are non-negative. A log-transformation is applied to all five, which substantially reduces skewness of the distributions. Covariates that have 0 values are shifted up by the value 1 before taking log, and thus non-negativity is preserved.%
}

{%
Next, each log($t$) is centered to reduce dependence among the corresponding regression coefficients in the Bayesian inference. Thus, the covariate vector is $\bs{x}_{ij}=[1, t_{ij}^{p\ast}, t_{i1}^{s\ast}, t_{i2}^{s\ast}, t_{j1}^{r\ast}, t_{j2}^{r\ast}]'$ where a ``$\ast$'' denotes log-transformed and centered.%
}
\subsection{Bayesian hierarchical model for the Benguela data}\label{sec-model}
{%
For these data, only 196 out of the 29$\times$28 = 812 pairwise links are non-zero. The sets ${\cal S}_3$=${\cal I}_3$=$\emptyset$ as there are no receive-only nodes. ${\cal S}_1$ consists of 5 pairs: $(i,j)$=(other groundfish, hakes), (other groundfish, squid), (hakes, squid), (birds, seals), and (seals, sharks). The cardinality of ${\cal S}_2, {\cal I}_1,$ and ${\cal I}_2$ are, respectively, 186, 2, and 27. Thus, given covariate vector $\bs{x}_{ij}$ for all $(i,j)$$\in$${\cal S}$, the model is:%
}%
\begin{gather}
\mu_{ij}(\bs{\beta})=\bs{x}_{ij}'\bs{\beta}\hspace{1cm}
\text{for all }(i,j)\in{\cal S}\,, \label{mu} \\
\begin{split}
\left.
[y_{ij},y_{ji}]' \right|
\bs{\beta},s_i,s_j,r_i,r_j,\bs{u}_i,\bs{u}_j,\bs{v}_i,\bs{v}_j,\bs{\Omega} \hspace{2in} \\
\sim\text{BVN}(
\left[
\begin{matrix}
\mu_{ij}(\bs{\beta}) \\
\mu_{ji}(\bs{\beta})
\end{matrix}
\right] +
\left[
\begin{matrix}
s_i \\
r_i
\end{matrix}
\right] +
\left[
\begin{matrix}
r_j \\
s_j
\end{matrix}
\right] +
\left[
\begin{matrix}
\bs{u}_i'\bs{v}_j \\
\bs{u}_j'\bs{v}_i
\end{matrix}
\right],
\bs{\Omega})
\hspace{1cm}\text{for all }(i,j)\in{\cal S}_1\,,
\end{split}\label{y-mutual} \\
y_{ij}\,|\,\bs{\beta},s_i,r_j,\bs{u}_i,\bs{v}_j,\sigma^2\sim
\text{N}\left(\mu_{ij}(\bs{\beta})+s_i+r_j+\bs{u}_i'\bs{v}_j,\sigma^2\right)
\text{\hspace{1cm}for all }(i,j)\in{\cal S}_2\,, \label{S2} \\
s_i|\sigma_s^2\sim\text{N}(0,\sigma_s^2)\,,\hspace{5mm}
r_i|\sigma_r^2\equiv -4\sigma_r \hspace{1cm}
\text{for all }i\in{\cal I}_1\,, \\
\left.
\left[
\begin{matrix}
s_i\\
r_i
\end{matrix}
\right]
\right|
\bs{\Sigma}
\sim\text{BVN}({\bf 0},\bs{\Sigma})
\text{\hspace{1cm}for all }i\in{\cal I}_2\,, \label{sr-mutual} \\
u_{iq}|\sigma_{uq}^2\sim\text{N}(0,\sigma_{uq}^2)\hspace{1cm}
\text{for all }i\in{\cal I}\text{ and }q=1,\ldots,k\,, \\
v_{iq}|\sigma_{vq}^2\sim\text{N}(0,\sigma_{vq}^2)\hspace{1cm}
\text{for all }i\in{\cal I}_2\text{ and }q=1,\ldots,k\,, \label{v.iq}
\end{gather}
{%
where $\bs{\beta}$ is the unknown regression coefficient.%
}

{%
Note that $\bs{u}_i,\bs{v}_j, \sigma_{uq}^2$, and $\sigma_{vq}^2$ are intrinsically unidentifiable (see \citep{pnas}), but $\bs{u}_i'\bs{v}_j$ is. To facilitate proper interpretation of feeding preference through the latent $\bs{u}$- and $\bs{v}$-spaces, the Procrustes transformation derived in \citep{pnas} can be applied to the Markov chain Monte Carlo (MCMC) scans~of~$\bs{u}_i$ and $\bs{v}_j$ for $k$=2. In the case of $k$=1, a constraint is built into the model to ensure identifiability (see \ref{sec-proc}).%
}

{%
For our analyses, we employ reasonably diffuse priors, coupled with various reparametrizations to mitigate MCMC mixing issues. See \ref{sec-prior.mix}.%
}

\subsection{St.~Martin data and model}\label{sec-model2}
{%
\citet{goldwasser} collated the St.~Martin diet data from various existing sources including their own, then subsequently introduced adjustments to the existing values as they saw fit. Their finalized diet data correspond to non-zero consumption frequencies in acts per hectare per day. Aside from brief descriptions of the 44 trophic species, no other associated data are available from \citep{goldwasser}. As we did for Benguela, here we created the pair-specific taxonomic distance according to the provided trophic species descriptions, then log-transformed and centered it to yield $\bs{x}_{ij}=[1,t_{ij}^{p\ast}]'$. We took $y_{ij}$ to be the 15th-root transformation of the weighted diet data in \citep{goldwasser} to achieve approximate normality of positive frequencies.%
}

{%
Of the 44 St.~Martin nodes, 6 are basal (${\cal I}_1$), 8 are top predators (${\cal I}_3$), and the rest are in between (${\cal I}_2$). Only 218 of all 1892 pairs exhibit feeding activity, of which 24 are send-only pairs (${\cal S}_2$) and 194 are receive-only (${\cal S}_3$). There are no mutually predatory pairs (${\cal S}_1=\emptyset$). Our model for the reduced St.~Martin web thus comprises (\ref{mu}), (\ref{S2})--(\ref{v.iq}), as well as%
}%
\par\noindent%
\begin{gather*}
y_{ji}\,|\,\bs{\beta},s_j,r_i,\bs{u}_j,\bs{v}_i,\sigma^2\sim
\text{N}\left(\mu_{ji}(\bs{\beta})+s_j+r_i+\bs{u}_j'\bs{v}_i,\sigma^2\right)
\text{\hspace{1cm}for all }(i,j)\in{\cal S}_3\,, \\
s_i|\sigma_s^2\equiv -4\sigma_s\,,\hspace{5mm}
r_i|\sigma_r^2\sim\text{N}(0,\sigma_r^2) \hspace{1cm}
\text{for all }i\in{\cal I}_3\,.
\end{gather*}
{%
See \ref{sec-prior.mix} for our choice of prior distributions.%
}
\subsection{MCMC implementation}
{%
Unlike \citep{pnas} which employed readily available MCMC software, we implemented the above models in WinBUGS \citep{winbugs} (version 1.4.3) and OpenBUGS \citep{openbugs} (various versions since 3.0.3). The actual choice of BUGS and version employed depended on the computer on which the data were analyzed.%
}
\section{Results}\label{sec-results}
\subsection{Benguela models for population consumption with all covariates}\label{sec-pop-5cov}
\begin{figure}[!h]
\begin{center}
\caption{Benguela results from the reduced model in Section \ref{sec-model}, assuming 2-dimensional $\bs{u}_i$ and $\bs{v}_j$ ($k$=2) and taking $y_{ij}$ as population consumption defined in (\ref{20th}). All plots are based on one of two MCMC chains. Left panels are Graphs U, and right panels, Graphs V. Top row: Graphs U and V showing the Benguela nodes (arrows point from prey to predator) with node label $i$ positioned at the posterior mean of $\bs{u}_i$ (left) and $\bs{v}_j$ (right). Bottom left: Graph U showing posterior draws of $\bs{u}_i$ for $i$=1, 15, and 16 only. Bottom right: Graph V showing posterior draws of $\bs{v}_j$ for $j$=16 and 29 only. \label{uv}}
\makebox{
\includegraphics[scale=.4]{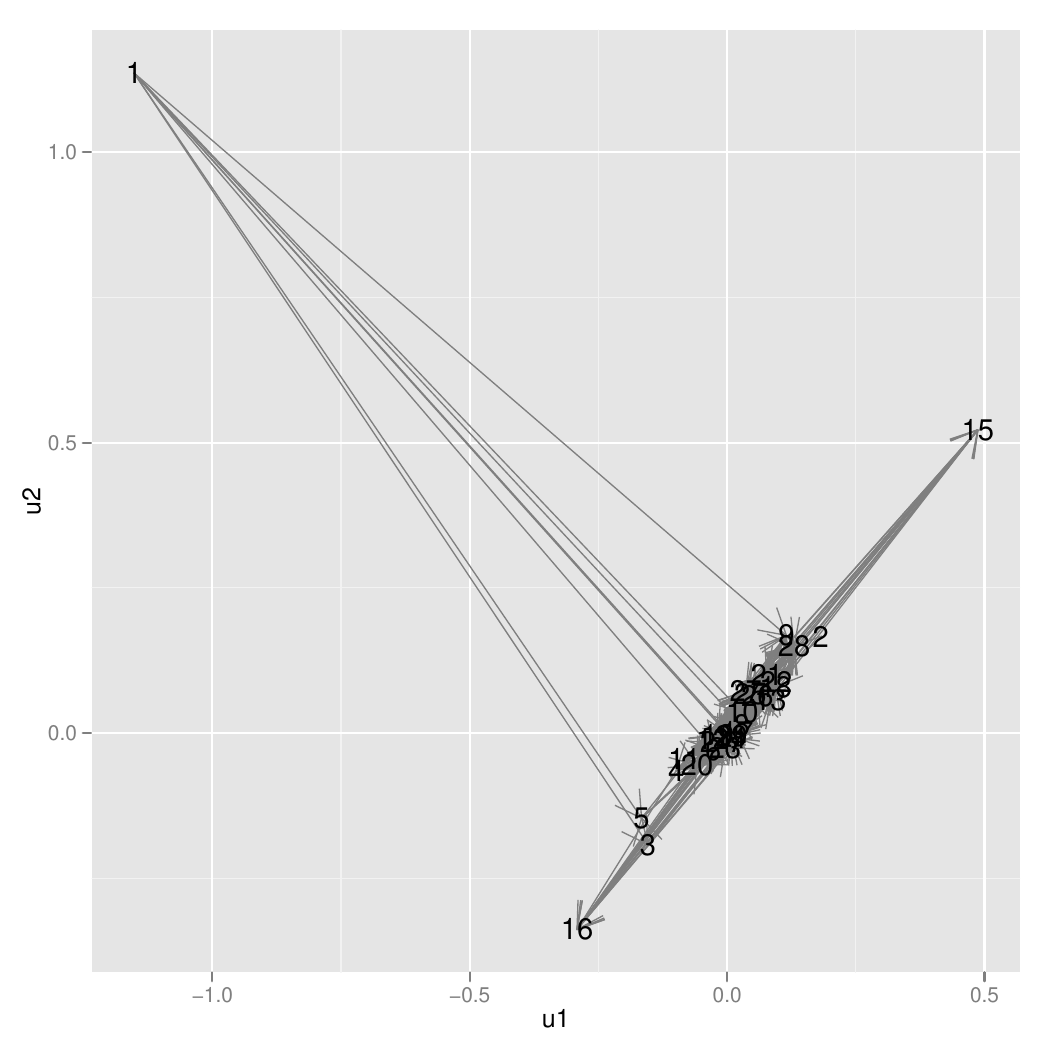}
\includegraphics[scale=.4]{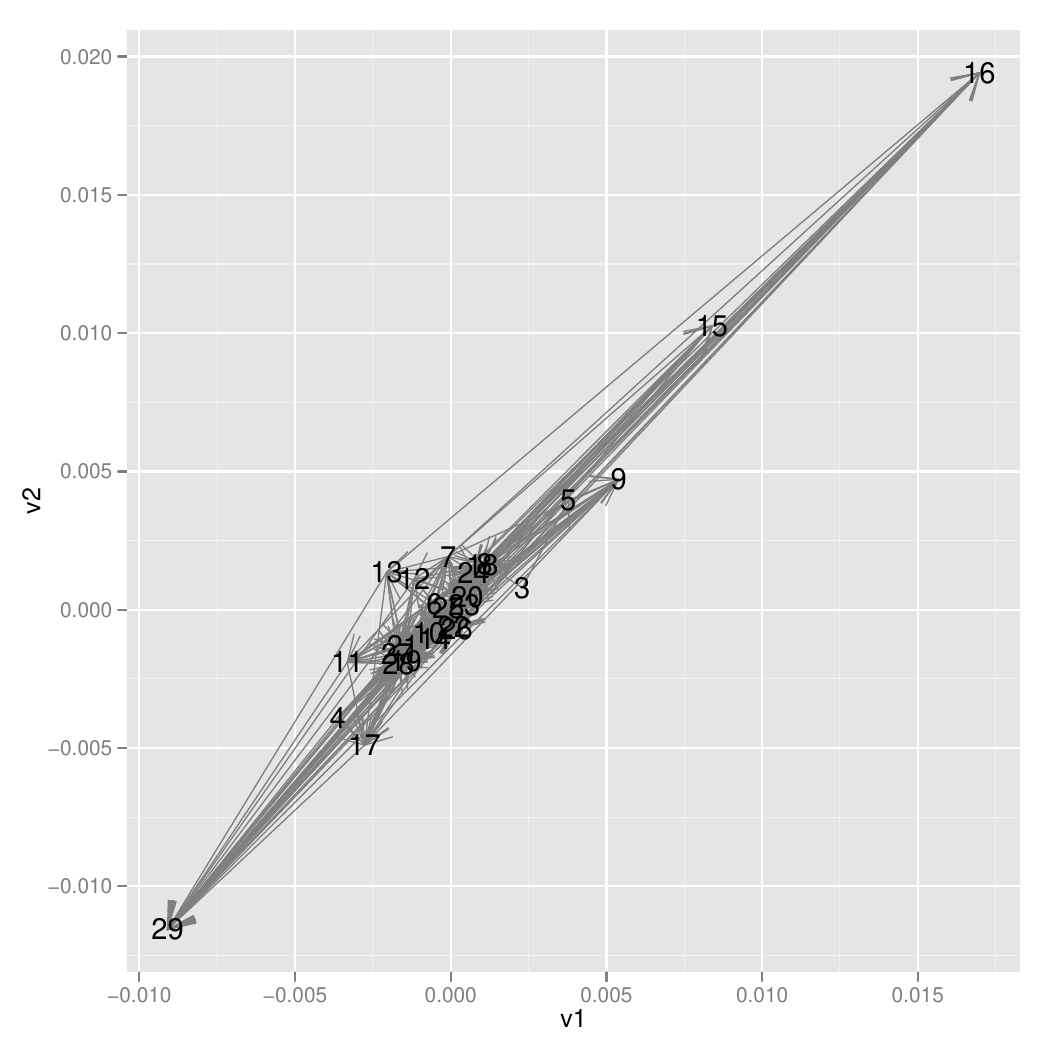}
}
\makebox{
\includegraphics[scale=.4]{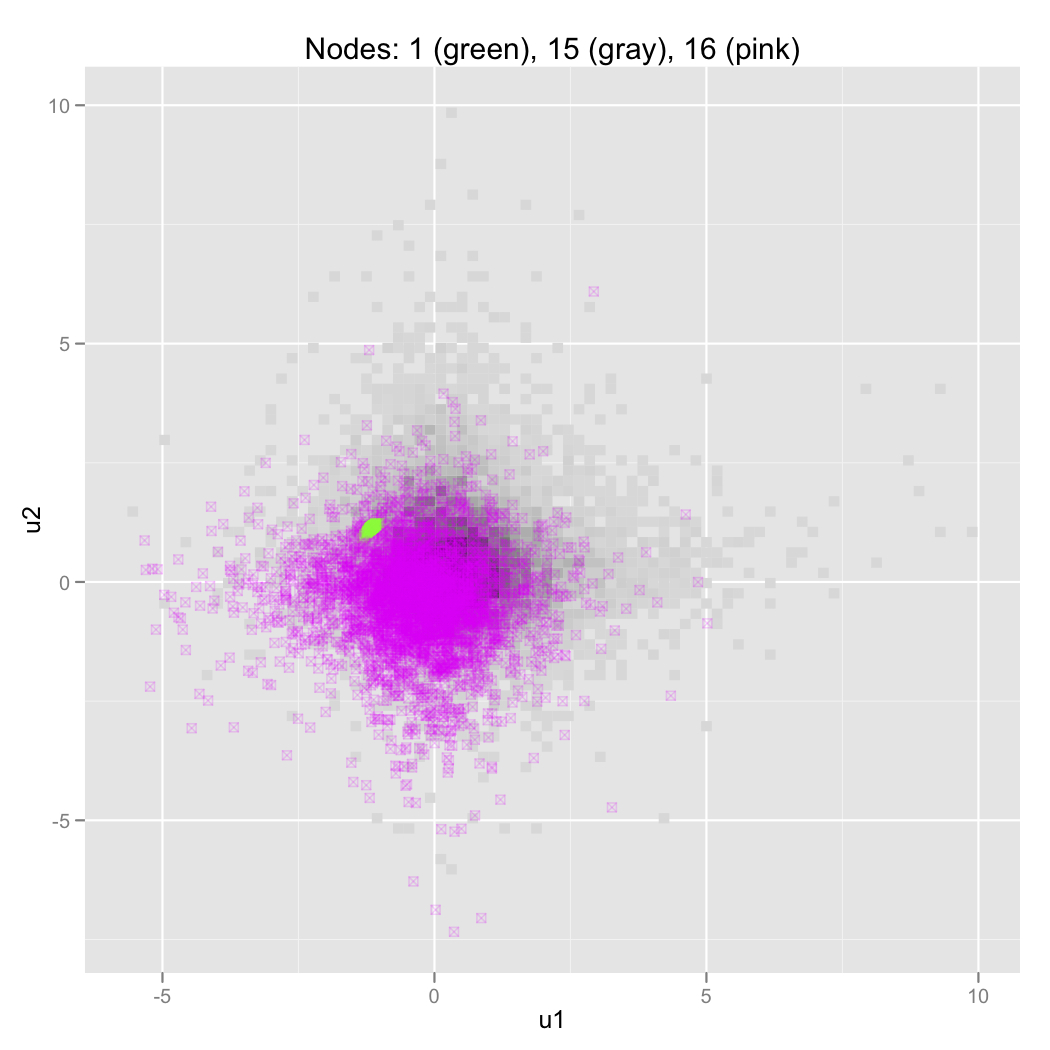}
\includegraphics[scale=.4]{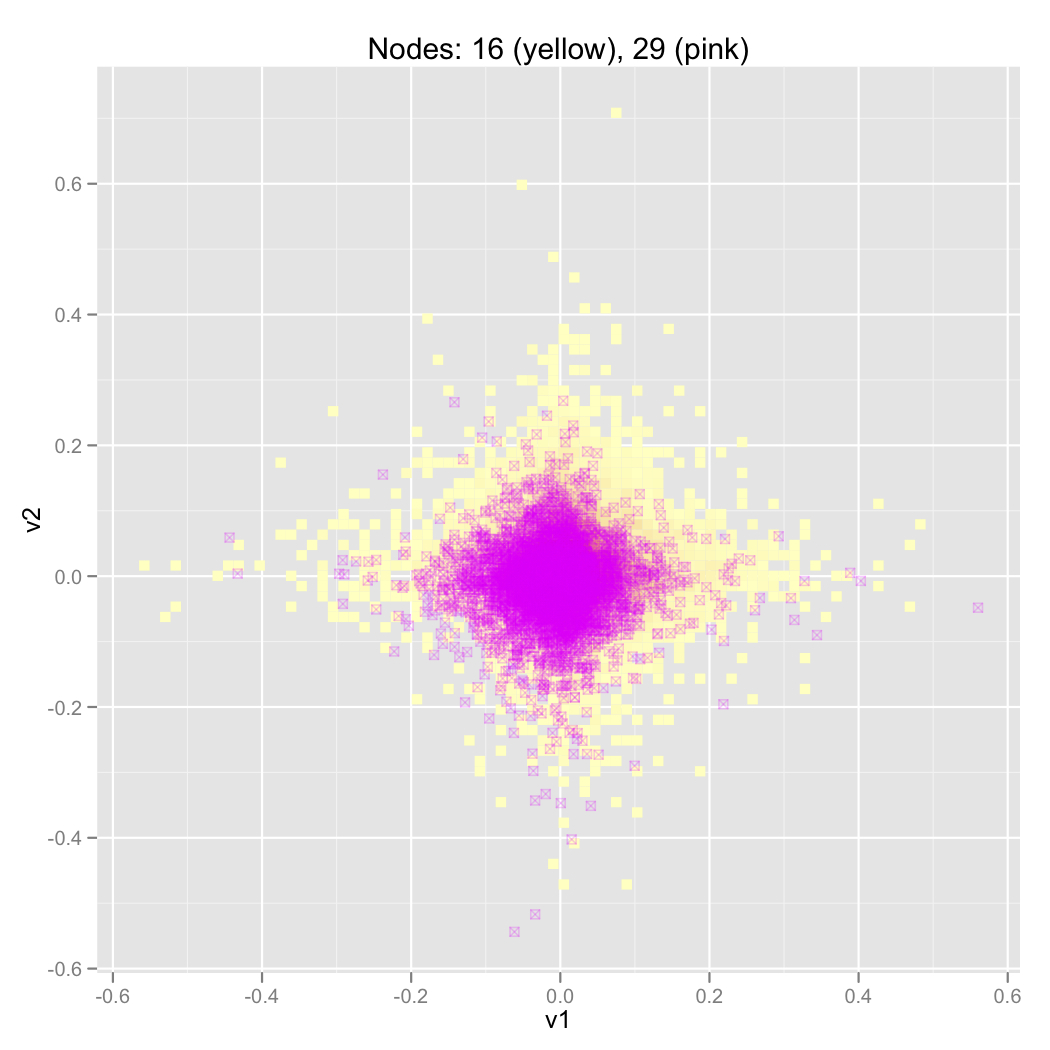}
}
\end{center}
\end{figure}

{%
We first applied the reduced model as described in Section \ref{sec-model}, with the population consumption defined in (\ref{20th}) as the response on all five covariates, taking $k$=2. Two MCMC chains were generated from distinct starting values, and convergence was monitored for each fixed and random parameter including $\bs{u}_i'\bs{v}_j$ (but excluding $\bs{u}_i,\bs{v}_j, \sigma_{uq}^2$, and $\sigma_{vq}^2$), and was deemed satisfactory. We observed that each of $\sigma_{uq}^2,\sigma_{vq}^2,\bs{u}_i$, and $\bs{v}_j$ for various $i$ and $j$ differed in location and spread between the two chains even after more than four million iterations, although their unidentifiability is irrelevant to proper inference given the convergence of $\bs{u}_i'\bs{v}_j$ for all $i,j$. As such, we arbitrarily selected the first chain, and subsequently applied the Procrustes transformation in \citep{pnas} to $\bs{u}_i$ and $\bs{v}_j$ to facilitate the interpretation of the preference spaces presented as Graphs U and V in Fig.~\ref{uv}. In the top left panel, the posterior mean for $\bs{u}_1$ (sending preference of phytoplankton) appears to differ drastically from the rest of the web. Aside from $i$=1, sending preference appears to differ much between $i$=15 (other pelagics) and 16 (horse mackerel). Yet, when posterior uncertainty is considered, the bottom left panel indicates that all three nodes are statistically indistinguishable from each other in the $\bs{u}$-space, although $\bs{u}_1$ is non-zero with very high credibility. Similar arguments apply to receiving preference (right panels), where posterior distributions overlap greatly between $j$=16 and 29 (sharks) despite their means being farthest apart from each other.
The extent of variability we observe here is very unlike that in \citep{pnas} with a logistic model for binary data including all 0s. Here, we contend that when only 24\% of all 812 links are modeled, inference is inadequate to distinguish among preference random effects, given a high-dimensional parameter space corresponding to all random effects and fixed parameters in the model.%
}%

\begin{figure}[!h]
\caption{Benguela results from fitting the reduced model in Section \ref{sec-model}, assuming $k$=1 and taking $y_{ij}$ as population consumption (\ref{20th}). Top panels: results under the 5-covariate model. Top left: Graph SR showing posterior means of $[s_i,r_i]'$ and corresponding MCMC samples for selected $i$. Top middle and right: Graphs U and V showing 80\% HPD intervals. Note that the interval for $u_1$ is positive (99\% HPD interval is [0.0003, 0.0009]), and that inference is inapplicable to $u_3,v_1$, and $v_2$. Bottom: same as top left but with predator harvest as the sole covariate and omitting posterior samples. \label{fig-sr.links}}
\begin{picture}(0,0)
\put(0,-250){
\includegraphics[height=3.26in]{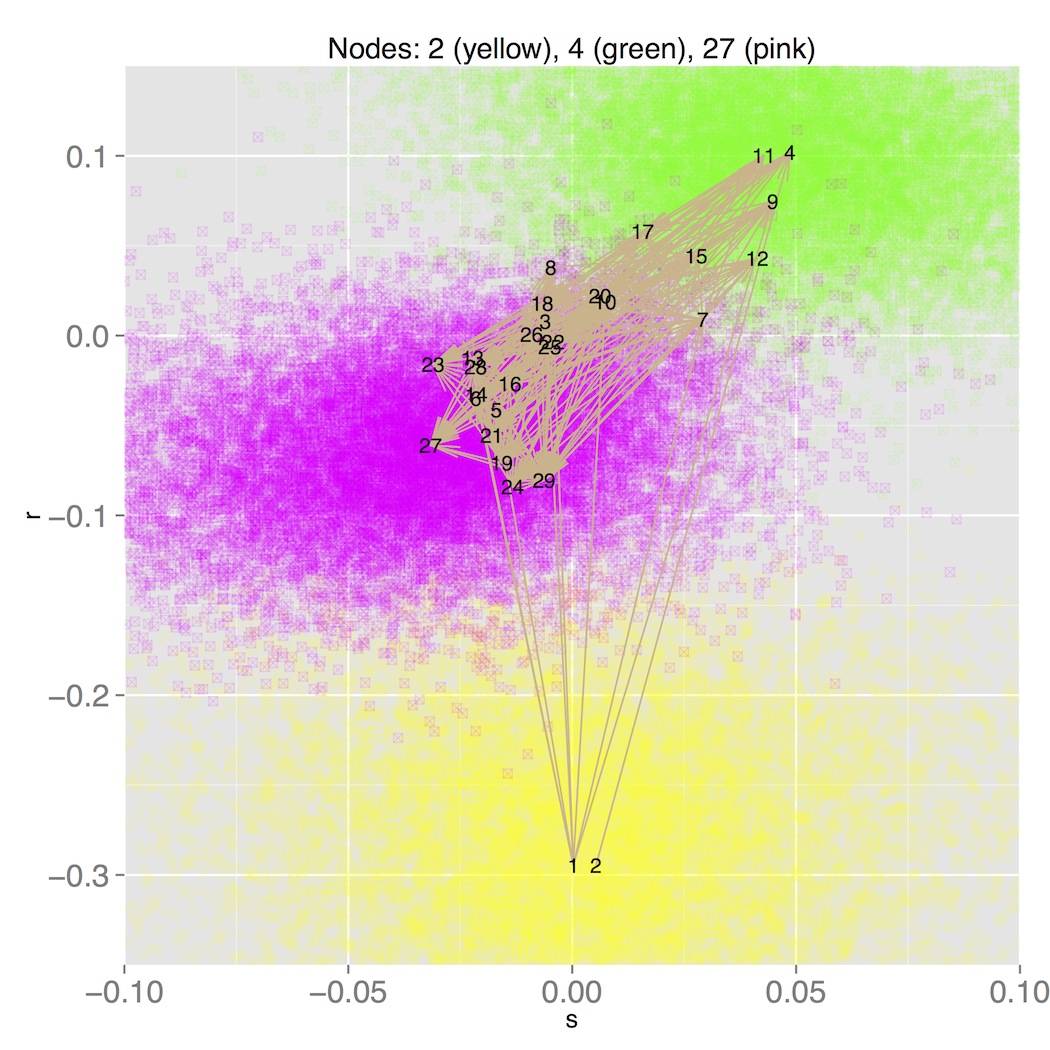}
}
\put(240,-260){
\includegraphics[height=3.27in]{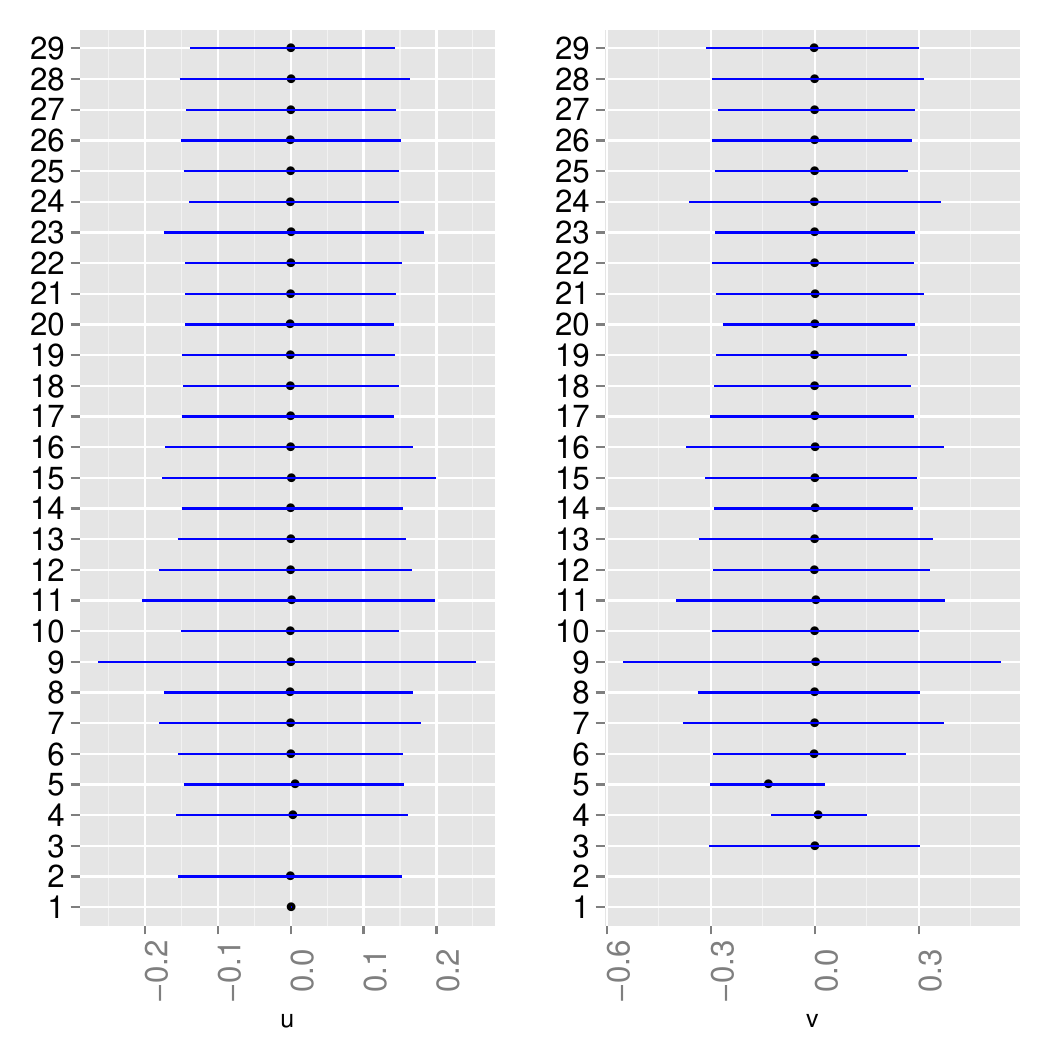}
}
\put(0,-510){
\includegraphics[height=3.5in]{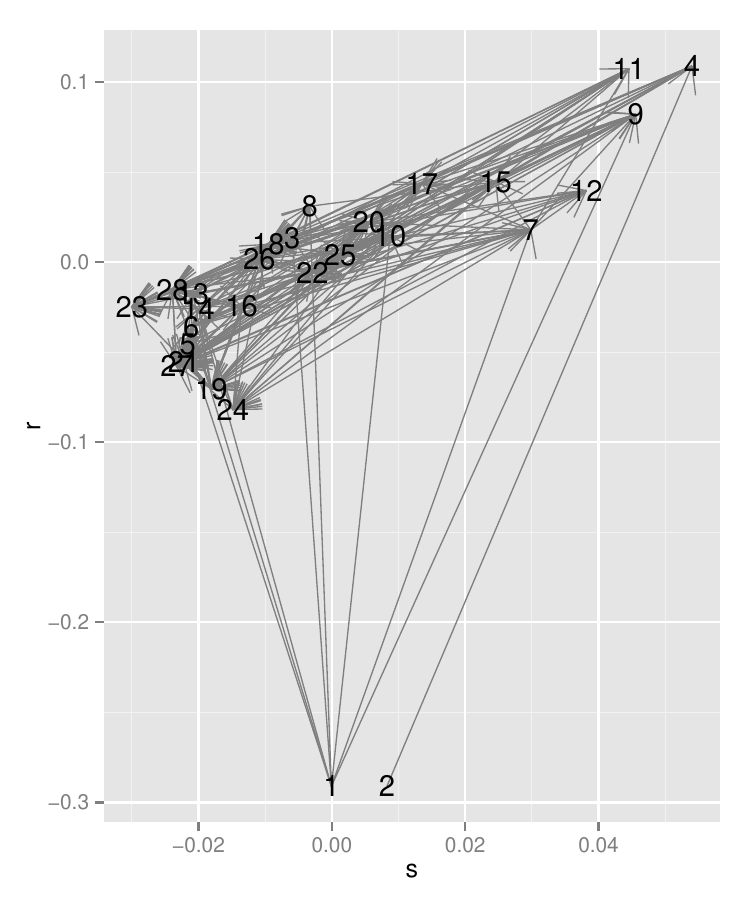}
}
\end{picture}
\end{figure}

{%
Note that the nearly linear alignment of posterior means (except for $i$=1 in the $\bs{u}$-space) and substantial posterior uncertainty seen in Fig.~\ref{uv} suggest that $k$=2 might be unnecessarily large. For this reason, we re-fitted the model with $k$=1; see \ref{sec-proc}. (All subsequent models for Benguela in this paper consider $k$=1.) MCMC convergence was quick. Results are shown in the top panels of Fig.~\ref{fig-sr.links}. Each node is placed at the posterior mean for $[s_i,r_i]'$ (feeding activity while adjusting for covariates). MCMC samples for $[s_i,r_i]'$ are also shown for $i$=2 (benthic filter-feeders), 4 (benthic carnivores), and 27 (birds). Although not shown for $i$=1 (phytoplankton), 9 (anchovy), or 11 (round herring), the posterior distributions are in fact very similar between the basal nodes $i\in$ \{1, 2\}, and among $i\in$ \{4, 9, 11\} which are middle nodes (${\cal I}_2$). Thus, the two basal nodes are clearly distinct from the rest of the food web. Overall, the graph indicates little overlap among the posterior distributions for 27 and the two sets \{1, 2\} and \{4, 9, 11\}. This suggests that the three show statistically distinguishable feeding activity with respect to non-zero consumption volume. It can be deduced that the posterior distributions for the remaining nodes fall between those for $i$=4 and 27, and may not be highly statistically distinguishable from each other. Distinguishability of the posterior distributions for feeding activity may be used to identify trophic clusters or trophic levels, which can facilitate subsequent steps in a typical food web analysis, such as the identification of keystone species. At present, we do not consider formal distance measures for multivariate distributions for this purpose. Inference for clustering in a special case of latent space models is considered by \citet{pavel1}; our current paper does not consider adapting their work to our case which concerns a reduced web with asymmetry between the $u$- and $v$-spaces.%
}

\begin{table}[!h]
\begin{center}
\caption{Benguela HPD intervals of interest for the reduced model in Section \ref{sec-model}, taking $u_i$s and $v_j$s as one-dimensional (see \ref{sec-proc}) and $y_{ij}$ as population consumption (\ref{20th}). Credible levels above 0.5 are presented in bold. \label{table-full-intervals}}
\begin{tabular}{lrrlr}\\\hline
&& \multicolumn{2}{c}{HPD interval} & \\ \cline{3-4}
Parameter & Posterior mean & Lower limit & Upper limit & Credible level \\ \hline
$\beta_1$ (taxo.~dist.) & 3.7$\times$10$^{-4}$ & \multicolumn{2}{c}{(interval includes 0)} & 0.50 \\
$\beta_2$ (prey biomass) & $-$1.2$\times$10$^{-3}$ & \multicolumn{2}{c}{(interval includes 0)} & 0.50 \\
$\beta_3$ (prey harvest) & $-$7.5$\times$10$^{-4}$ & \multicolumn{2}{c}{(interval includes 0)} & 0.50 \\
$\beta_4$ (pred.~biomass) & 2.0$\times$10$^{-3}$ & 0.000 & 0.004 & {\bf 0.60} \\
$\beta_5$ (pred.~harvest) & 5.5$\times$10$^{-4}$ & \multicolumn{2}{c}{(interval includes 0)} & 0.50 \\
$\rho_{sr}$ & 0.429 & 0.075 & 0.967 & {\bf 0.80} \\
$\rho$ & 0.328 & 0.070 & 0.802 & {\bf 0.75} \\ \hline
\end{tabular}
\end{center}
\end{table}

{%
Here, Graph SR for the reduced model shows that non-zero biomass sent and non-zero biomass received are somewhat positively correlated (disregarding $i=$1 and 2). This is also reflected by the posterior mean of $\rho_{sr}$ in Table \ref{table-full-intervals} (80\% highest probability density (HPD) interval is above 0). Table \ref{table-full-intervals} also indicates that among the 5 mutually consuming pairs in ${\cal S}_1$, the tendency of predator-prey role reversal is high ($\rho>$ 0) with strong evidence (80\% HPD interval has a lower limit of just below 0). Thus, in ${\cal S}_1$, predators who consume much (little) tend to be consumed heavily (limitedly) by volume themselves. Recall that the interpretation here corresponds to biomass (body size) and other covariates being adjusted for.%
}%

{%
We can also use Table \ref{table-full-intervals} to assess the statistical relevance of the five covariates in the social network model. We see that $\beta_4$ is the only regression slope parameter with more than a 0.5 posterior probability (credibility) for it to be in some interval that excludes 0. This suggests that predator biomass is reasonably relevant to explaining positive population consumption, while the other four covariates are essentially irrelevant. (Note that taxonomic distance is shown in \citep{pnas} to be relevant to the complete set of 0-1 data).%
}

{%
Finally, Graphs U and V in Fig.~\ref{fig-sr.links} suggest that reducing the preference space dimensionality $k$ from 2 to 1 does not increase the statistical distinguishability among the preference random effects. In contrast, the deviance information criterion (DIC) increased drastically from $-$297 to $-$237. The magnitude of the difference here suggests better model performance with $k$=2 instead of $k$=1.%
}
\subsection{A St.~Martin model}
{%
We digress from the Benguela case study to compare the above results with those of the St.~Martin web. Specifically, for the latter being a much larger web but with less than 12\% non-zero links, do the preference spaces show as much uncertainty as those for Benguela?%
}

\begin{figure}[!h]
\begin{center}
\caption{St.~Martin results from the reduced model in Section \ref{sec-model2}, assuming $k$=2. All plots are based on one of two MCMC chains. Top row: Graph U (left) and Graph V (right) with node labels at posterior means. Bottom left: Graph U showing posterior draws of $\bs{u}_i$ for $i$=18, 36, 40, 41, and 44 only.
Bottom right: Graph V showing posterior draws of $\bs{v}_j$ for $j$=2, 19, and 35 only.
\label{uv2}}
\makebox{
\includegraphics[scale=.45]{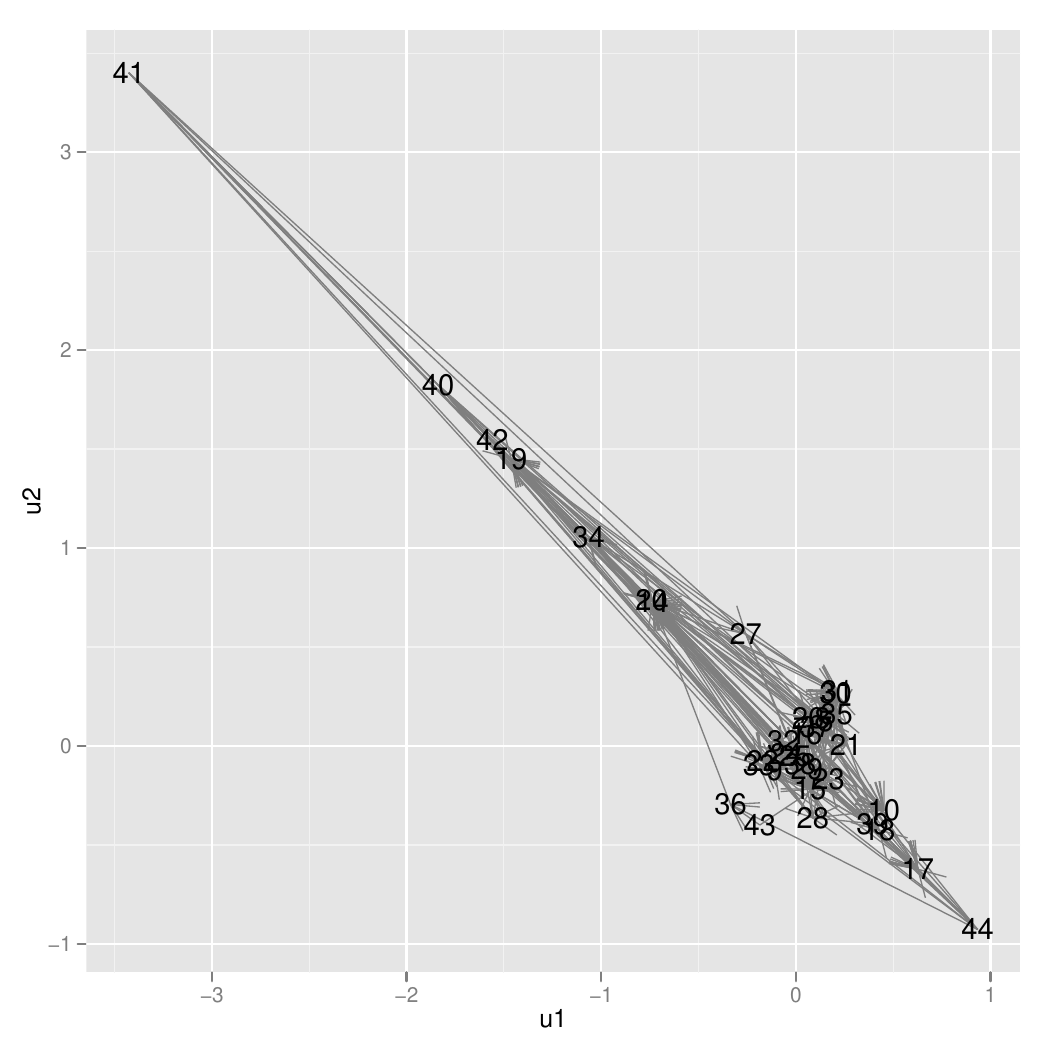}
\includegraphics[scale=.45]{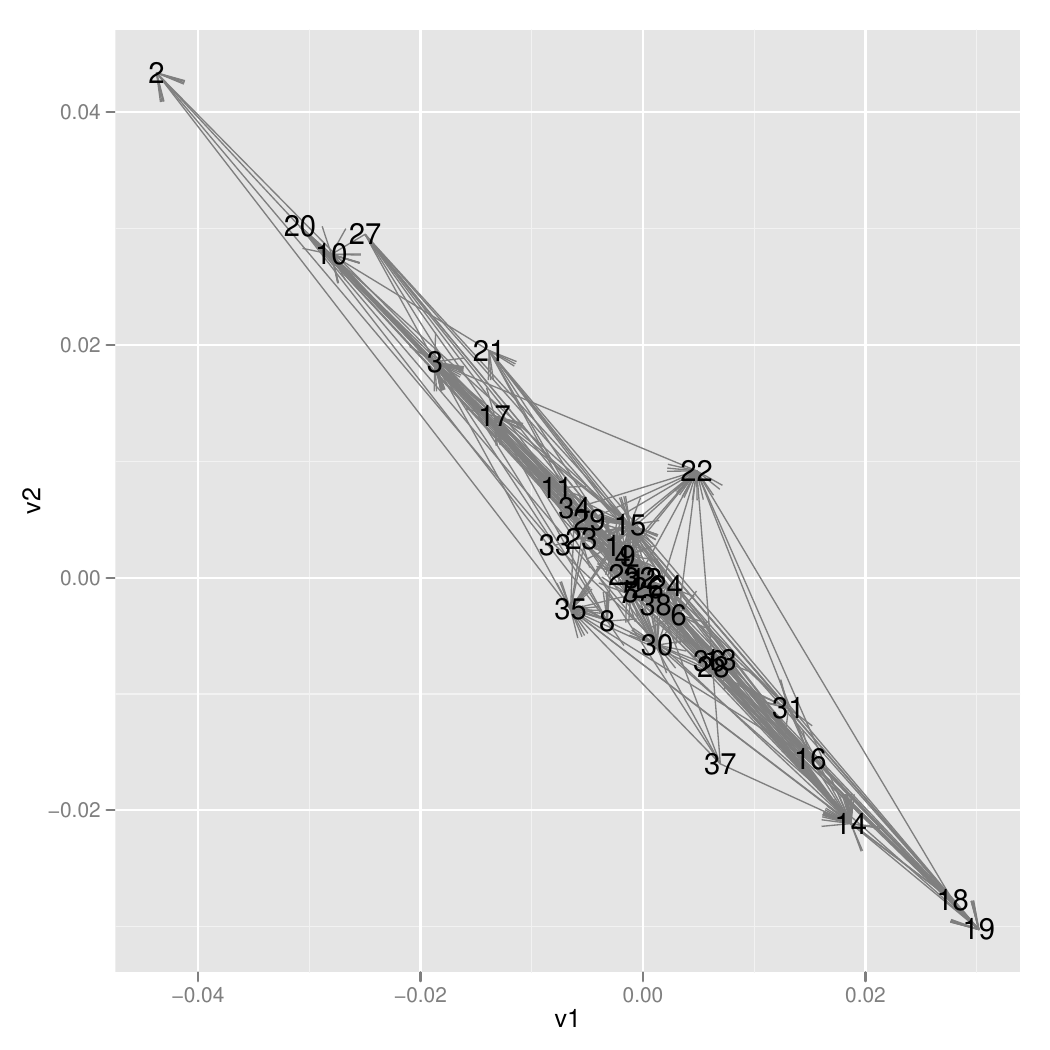}
}
\makebox{
\includegraphics[scale=.45]{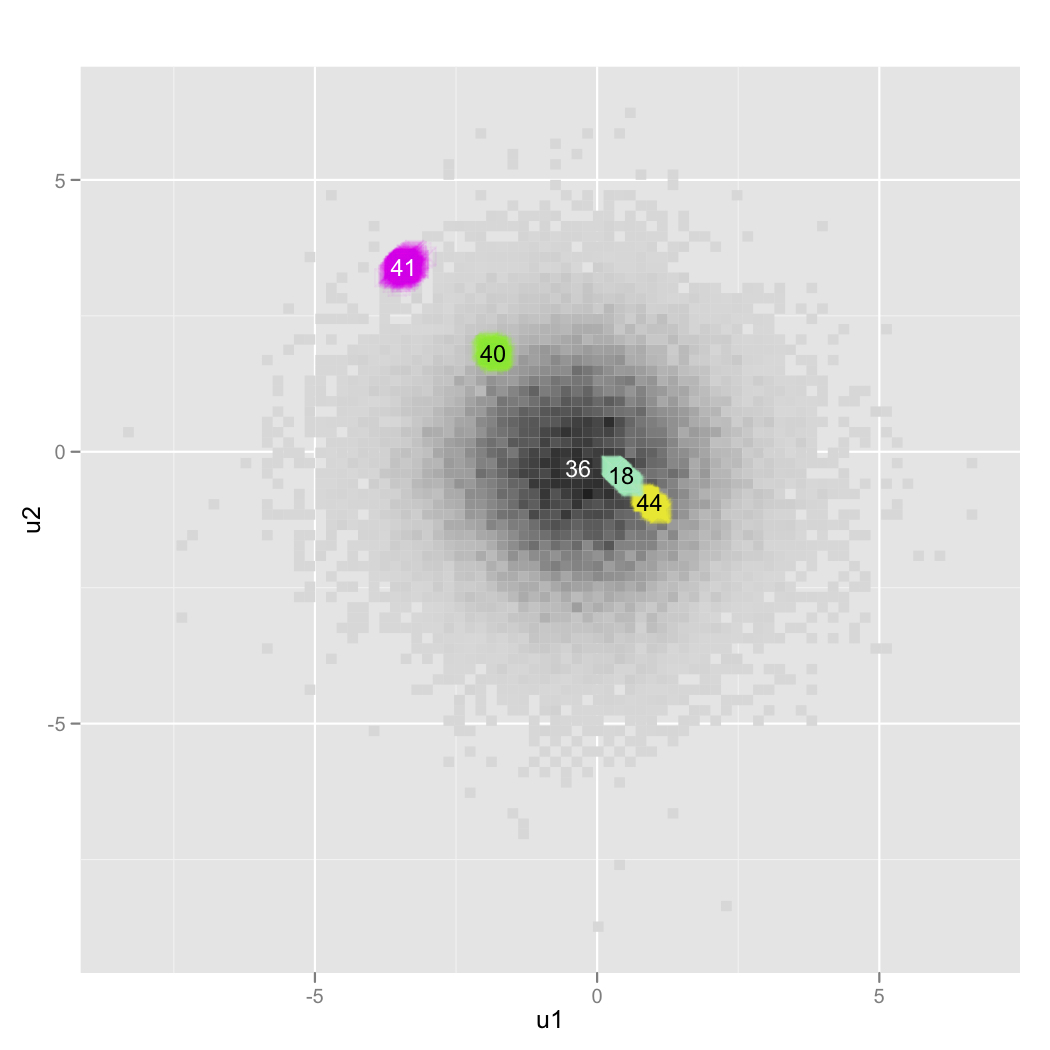}
\includegraphics[scale=.45]{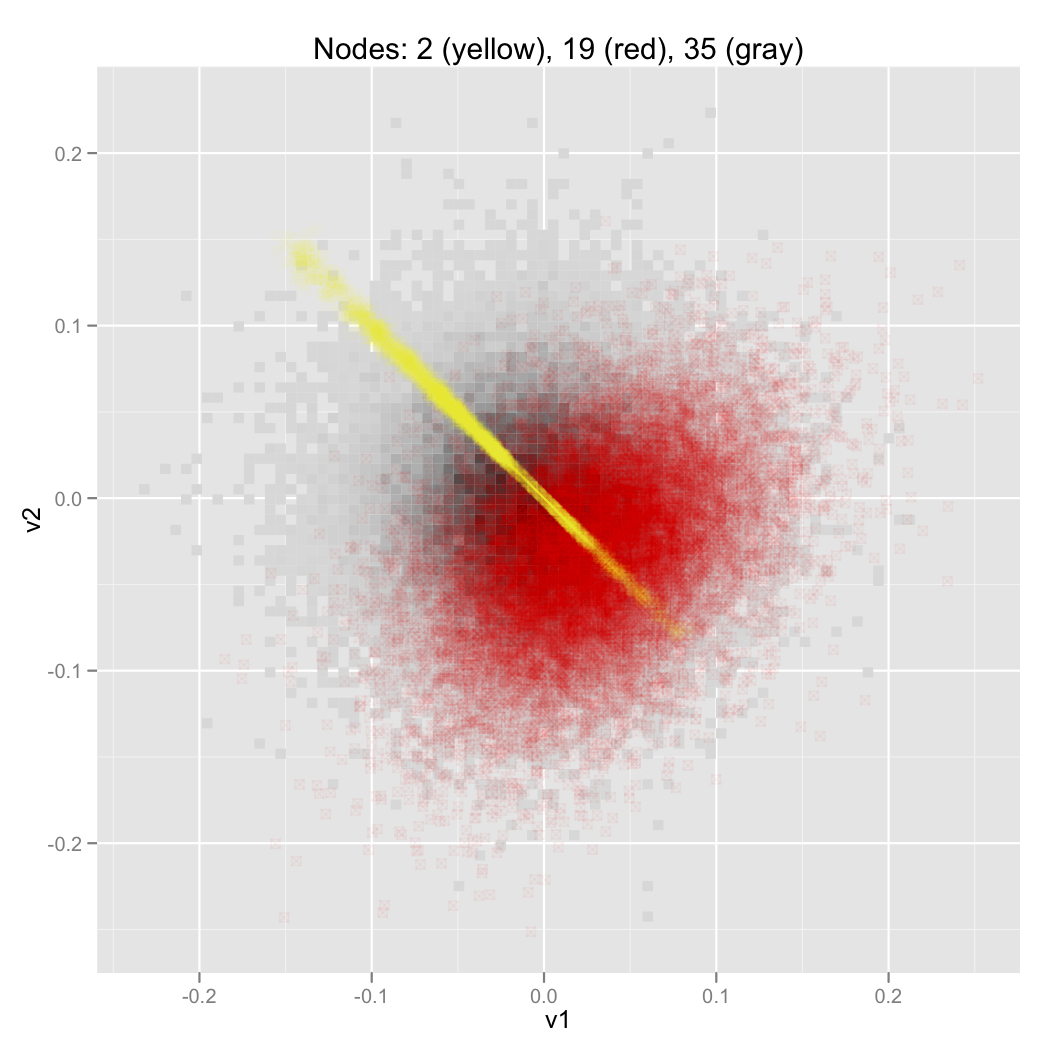}
}
\end{center}
\end{figure}

{%
For $k$=2, we used BUGS to generate multiple MCMC chains from distinct starting values, and convergence for all intrinsically identifiable parameters was satisfactory 
except for mixing issues for $\bs{u}_i'\bs{v}_j$ for $i$=25 (other hymenoptera) and $j$=14 ({\it A.~gingivinus}), 
and minor issues for a few other $(i,j)\in{\cal S}_2,{\cal S}_3$.
In the near future, we intend to explore novel numerical integration algorithms to address this fully, rather than relying on existing MCMC software. We now consider an exploratory data analysis by arbitrarily selecting one BUGS chain only (and therefore we do not report a DIC value). This roughly corresponds to constraining $\bs{u}_{25}'\bs{v}_{14}$ to a fixed value, given the nature of the various chains generated. We applied to this chain the Procrustes transformation for the preference spaces, and the resulting Graphs U and V (Fig.~\ref{uv2}) were similar to the Benguela case with $k$=2, i.e.,~very large posterior uncertainties for most nodes and a nearly linear alignment of the preference posterior means. However, unlike Benguela, various St.~Martin nodes appeared highly distinguishable among each other in their sending preferences (e.g., among nodes 18: isoptera, 40: nectar and floral, 41: leaves, and 44: detritus, and between nodes 36: millipede and 41). Moreoever, at least one node (2: scaly-breasted thrasher) showed noticeably less posterior uncertainty in its receiving preference than other nodes (although posterior distributions overlapped substantially among $j$=2 and 19: hemiptera, whose posterior means for $\bs{v}_j$ were farthest apart). One can deduce from Fig.~\ref{uv2} that all other nodes overlapped with the very large uncertainty cloud of $i$=36 in the $\bs{u}$-space, and of $j$=35 (centipede) in the $\bs{v}$-space. Here, small posterior spread is not restricted to basal or top predatory nodes, as demonstrated by the middle node $i$=18.
}%

\begin{table}[!ht]
\begin{center}
\caption{St.~Martin HPD intervals of interest for the reduced model in Section \ref{sec-model2}, with $k$=1. \label{table-martin-intervals}}
\begin{tabular}{lrrrr}\\\hline
&& \multicolumn{2}{c}{HPD interval} & \\ \cline{3-4}
Parameter & Posterior mean & Lower limit & Upper limit & Credible level \\ \hline
$\beta_1$ (taxo.~dist.) & 0.296 & 0.148 & 0.442 & {\bf 0.99} \\
$\rho_{sr}$ & $-$0.524 & $-$0.896 & $-$0.039 & {\bf 0.95} \\ \hline
\end{tabular}
\end{center}
\end{table}

\begin{figure}[!h]
\begin{center}
\caption{Graph U (left panels) and Graph V (right panels) for St.~Martin, showing 80\% HPD intervals based on the reduced model in Section \ref{sec-model2} with $k$=1. Bottom panels: replotting of very short intervals from the top panels. Note that inference is inapplicable to $u_1,\ldots,u_6,u_{11},u_{12},u_{13},v_{39},\ldots,v_{44}$. \label{uv2-1d}}
\includegraphics[height=4in]{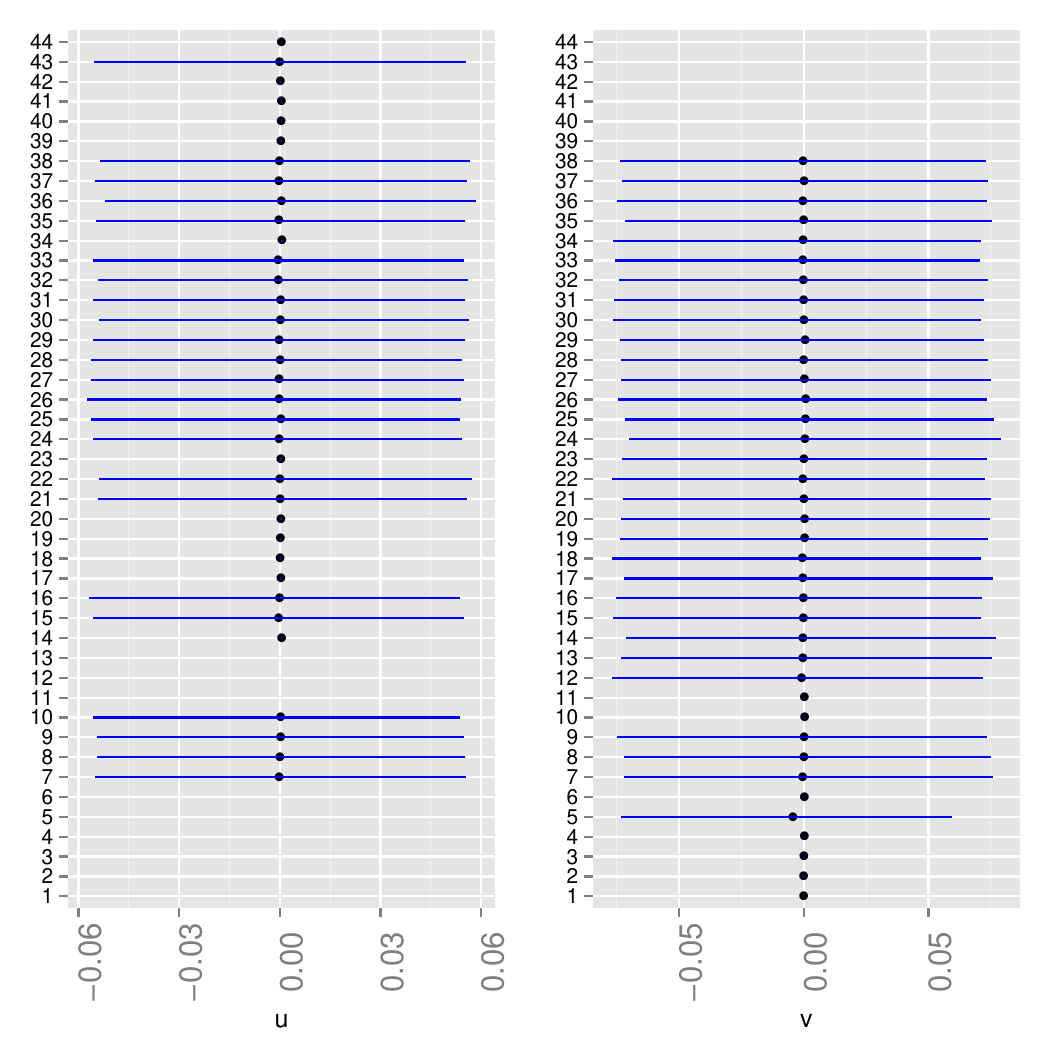}\\
\includegraphics[height=4in]{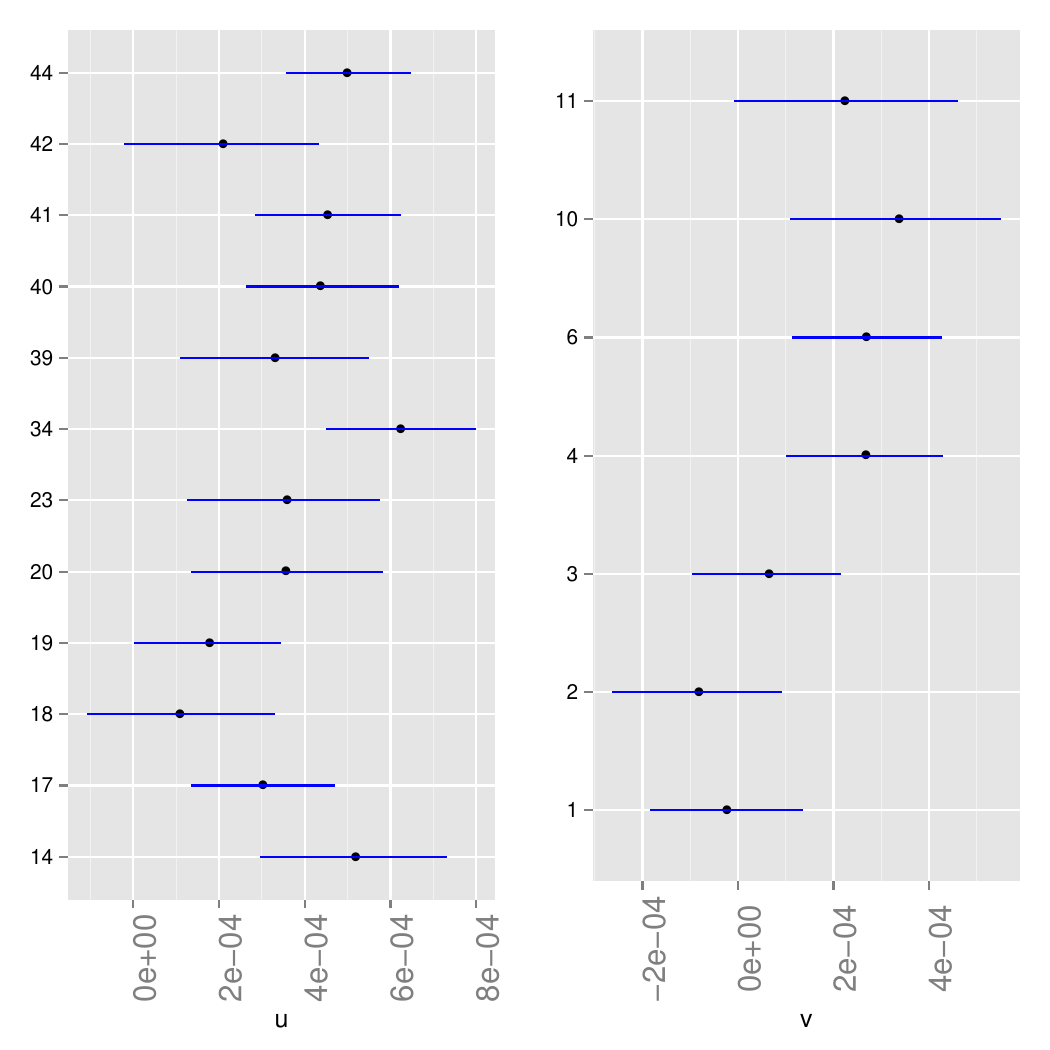}
\end{center}
\end{figure}

{%
Again, given the nearly linear alignment of nodes, we re-fitted the model but taking $k$=1 (Fig.~\ref{uv2-1d}). Overall conclusions about node distinction are similar regardless of $k$, except that certain pairs of nodes may be distinguishable under $k$=2 but not $k$=1 (e.g., $i$=41 and 44 in sending preference). This suggests that some biological information could be lost by collapsing the preference space dimensionality.%
}

{%
With respect to covariates, unlike Benguela, there is strong evidence for St.~Martin that taxonomic distance contributes significantly to feeding behavior: the larger the distance, the higher the consumption frequency. This can be seen in Table~\ref{table-martin-intervals}, where the positive HPD interval for $\beta_1$ has very high credibility (99\%). Also unlike Benguela is the high credibility (95\%) for the negative HPD interval for $\rho_{sr}$. Thus, active senders by volume tend to be inactive receivers by volume, and {\it vice versa}, even while adjusting for taxonomic distance.%
}
\subsection{The question of seal cull in the Benguela ecosystem}\label{sec-cull}
{%
We return to focus on the Benguela case study.%
}

{%
A large parameter space could be the reason for the weak credibility of non-zero regression coefficients seen in Section \ref{sec-pop-5cov}. To investigate this, we re-fitted the model with various subsets of the covariates, and report here the fit with a single covariate, predator harvest ($t_{j2}^{r\ast}$). This choice of covariate is due to the scientific interest of the potential benefits of a seal cull to future commercial harvest of fish species, including anchovy, horse mackerel, and hakes, upon which African fur seals predate \citep{yodzis}. Indeed, this single-covariate model has a noticeably smaller DIC (=$-$245) than the five-covariate model above (DIC=$-$237), suggesting a reasonable balance between fit and model parsimony.%
}

{%
The bottom panel of Fig.~\ref{fig-sr.links} shows Graph SR under this 1-covariate model. Visually, it is~comparable to that of the 5-covariate model (top left panel in the figure), but has a more noticeable positive trend among middle nodes (${\cal I}_2$). Indeed, HPD intervals for $\beta_5,\rho_{sr}$, and $\rho$ from this 1-covariate model (Table \ref{table-1cov-intervals}) all exclude 0 with reasonably high credibility, suggesting that the weak credibility for $\beta_5$ based on the larger model may well be due to insufficient data. Also, removing predator biomass (moderately high credibility earlier) and three other covariates from the model has increased the credible level of $\rho_{sr}$. This reflects that predator biomass is not only relevant to consumption volume, but also to the tendency in ${\cal I}_2$ to send and receive. Note that the high credibility of $\beta_5$ suggests a positive relationship between predator harvest and non-zero population consumption. The influence is noticeable in the plot of $y_{ij}$$>$0 against $t_{j2}^{r\ast}$ (Fig.~\ref{fig-y-vs-predharv}).%
}%
\begin{table}[t]
\begin{center}
\caption{Benguela HPD intervals of interest for the reduced model as for Table \ref{table-full-intervals}, except for taking predator harvest as the sole covariate.\label{table-1cov-intervals}}
\begin{tabular}{ccrlr}\\\hline
&& \multicolumn{2}{c}{HPD interval} & \\ \cline{3-4}
Parameter & Posterior mean & Lower limit & Upper limit & Credible level \\ \hline
$\beta_5$ & 0.003 & 0.001 & 0.006 & {\bf 0.95} \\
$\rho_{sr}$ & 0.488 & 0.083 & 0.975 & {\bf 0.85} \\
$\rho$ & 0.478 & 0.017 & 0.897 & {\bf 0.90} \\ \hline
\end{tabular}
\end{center}
\end{table}

{%
To investigate the relevance of predator harvest at the node level instead of the food web level, next we considered models with one random slope per sender node:%
\[
y_{ij}=\beta_0+\beta_{i5} t_{j2}^{r\ast}+s_i+r_j+\bs{u}_i'\bs{v}_j+\eps_{ij}\,,\ \ \ \ \ \ 
\beta_{i5}\,|\,\beta_5,\sigma_{\beta5}^2\sim\text{N}(\beta_5,\sigma_{\beta5}^2)\,.
\]
With additional slope parameters to account for randomness, the posteriors for preference random effects were extremely diffuse even with $k$=1 and priors for precision parameters that were somewhat restrictive (see \ref{sec-prior.mix}). This and DICs (Table \ref{dic}, last two rows) suggested that $\bs{u}_i'\bs{v}_j$ was no longer an important component in the model.%
}

{%
Thus, we summarize the fitted random-slope model with $\bs{u}_i'\bs{v}_j$ removed. HPD intervals for $\beta_{i5}$ appear in Table \ref{random-slope}. For this model, HPD intervals for $\beta_{i5}$ excluding 0 with a credibility of at least 70\% are observed for many non-basal sender nodes including anchovy and hakes ($i$=9 and 19, in red and blue, respectively, in Fig.~\ref{fig-y-vs-predharv}). Among these nodes, some have positive slopes and others have negative slopes. However, many other non-basal sender nodes (including $i$=16: horse mackerel, in green in the figure), as well as the overall slope $\beta_5$ correspond to a credibility of 50\% or less for intervals excluding 0. This contrasts with the high credibility for a positive $\beta_5$ in the fixed-slope model of Table \ref{table-1cov-intervals}. The discrepancy could be due to the extra variability imposed by the random-slope model, which may not be warranted for the Benguela data according to the larger DIC (weaker inference) as shown in Table \ref{dic}. Although the fixed-slope model with preference random effects has the smallest DIC among the 1-covariate models we investigated, the current random-slope model nonetheless provides scientific information about the relevance of predator harvest at the node level.
}

{%
In the causal context, a positive HPD interval (at either the food-web level or the node level) would be evidence that increasing the harvest from predator populations increases these predators' total consumption of prey, and hence, decreases the total availability of prey species. This reasoning is obviously counterintuitive as far as theorized population dynamics are concerned. Of course, our regression models are fitted to entirely empirical, observational data, so that causal conclusions are inapplicable. The positive association here between non-zero $y_{ij}$ and $t_{j2}^{r\ast}$ for a given prey $i$ (e.g., hakes) reflects the practice that humans tend to harvest more heavily those predator populations with a high consumption of biomass of prey $i$.%
}

{%
A more direct approach to address the potential influence of a seal cull on the food web without employing population dynamics would be to apply a longitudinal social network (LSN) model \citep{westveld} to a food web that is repeatedly observed over time. Although still observational and non-causal, a temporal model would incorporate any temporal fluctuations in the relationship between predator and prey. Given observations made at regular time intervals over a substantial time period, such fluctuations should reflect the underlying population dynamics. For example, time points at which the seal population biomass increases is expected to correspond (subject to natural variability) to the time points at which total consumption of prey species of seals also increases, and {\em vice versa}. Then, to assess the statistical significance of the influence of a seal cull (via increased seal harvest) on the availability of commercial fish species which are seals' prey, one may perform a posterior predictive analysis after fitting the LSN model. Any statistical significance would be entirely due to empirical information through the temporal data for consumption and covariates for the food web, and could serve as validation of projected phenomena based on population dynamics theory.%
}

{%
However, given the rarity of temporal food web diet data, an entirely empirical approach for validating population dynamics may be challenging. Proxy data based on stable isotopes \citep{jennings,pinnegar,winemiller} that are repeatedly collected over time do exist.
We intend to pursue LSN modeling of these temporal stable-isotope data in the future.%
}
\subsection{Benguela models for per-adult consumption}
\begin{figure}[!h]
\caption{Benguela results from fitting the reduced model in Section \ref{sec-model}, assuming 1-dimensional preference random effects $u_i$ and $v_j$ ($k$=1) and taking $y_{ij}$ as per-adult consumption (\ref{10th}). Top panels: results under the 5-covariate model. Top left: Graph SR, showing the Benguela food web (arrows point from prey to predator), with node label $i$ positioned at the posterior mean of $[s_i,r_i]'$, and corresponding MCMC samples shown for selected $i$. Top middle and right: Graph U and Graph V, respectively showing 80\% HPD intervals for $u_i$ and $v_j$. Note that the interval for $u_1$ is extremely short but includes 0, and that inference is inapplicable to $u_3,v_1$, and $v_2$. Bottom panel: same as top left but with no covariates and omitting posterior samples. \label{fig-sr-adult}}
\begin{picture}(0,0)
\put(0,-250){
\includegraphics[height=3.26in]{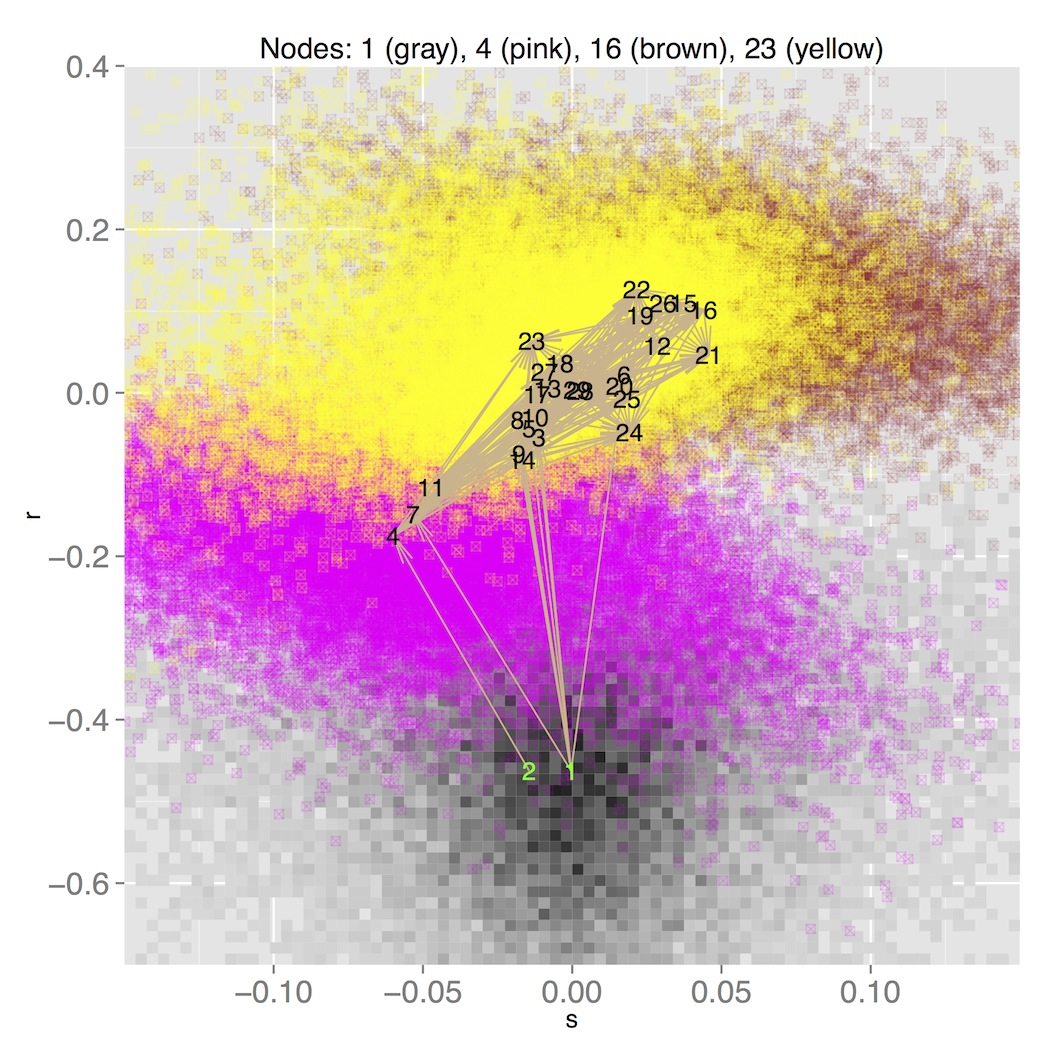}
}
\put(240,-260){
\includegraphics[height=3.27in]{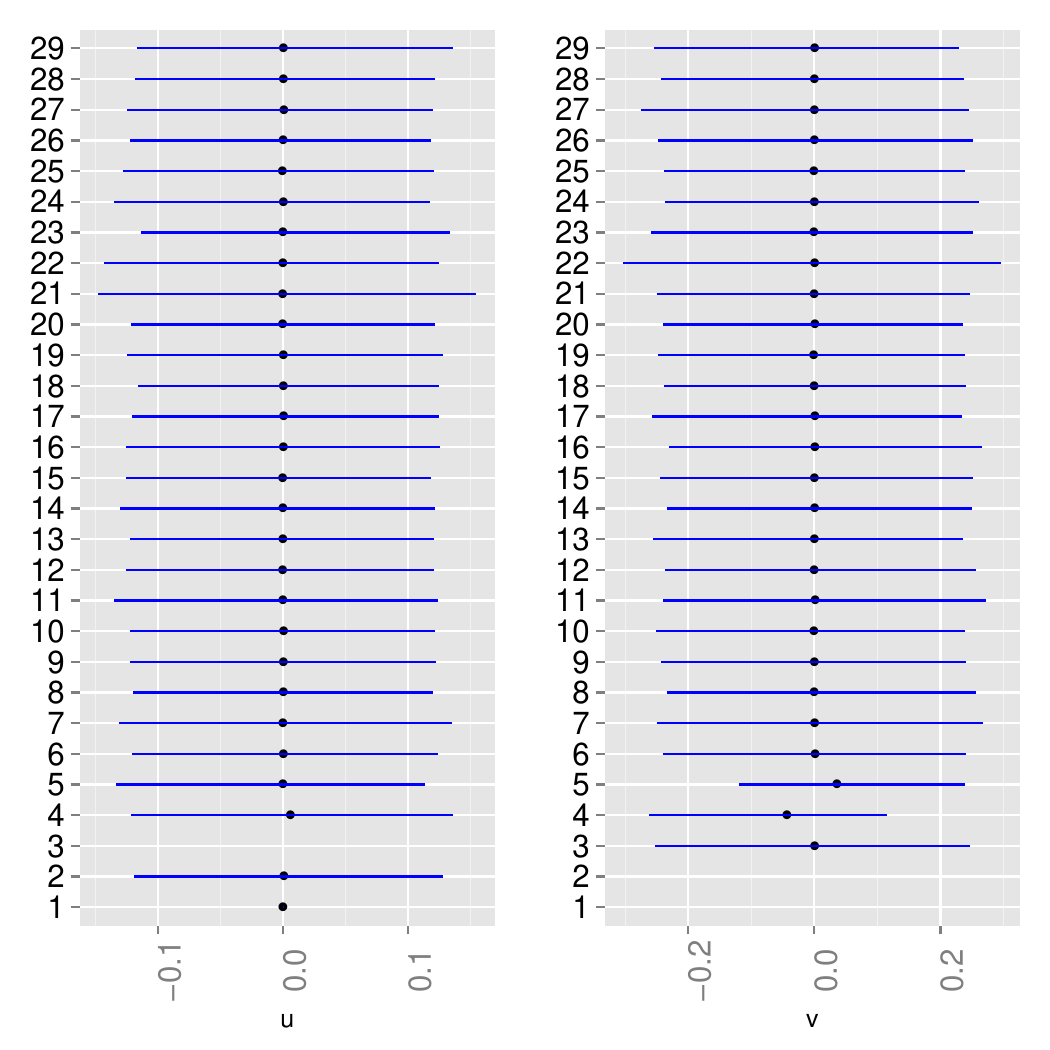}
}
\put(0,-510){
\includegraphics[height=3.5in]{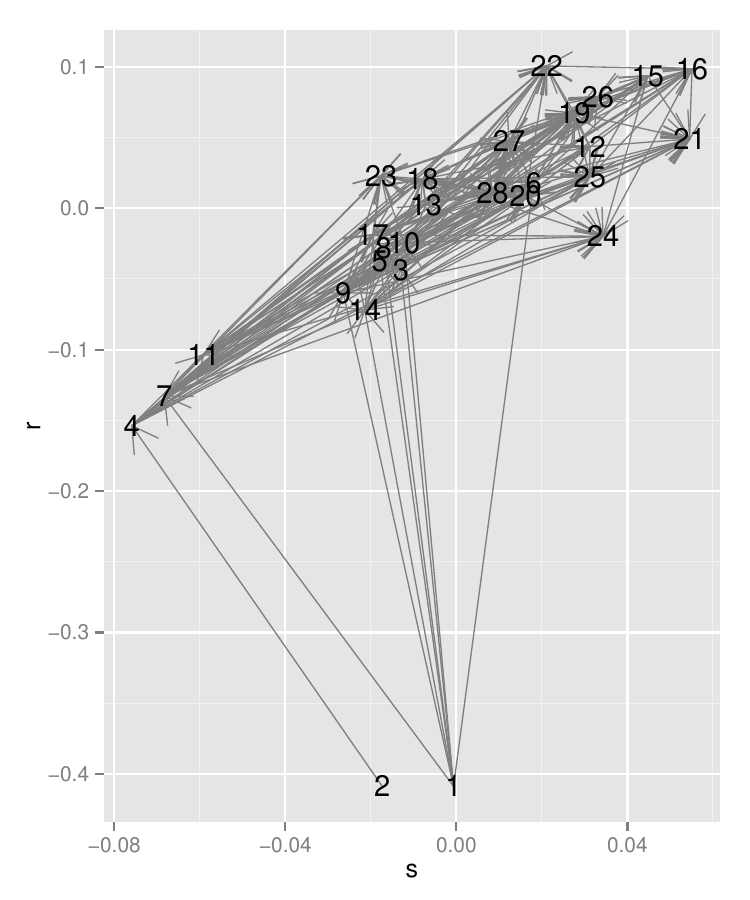}
}
\end{picture}
\end{figure}

{%
Although the above 5-covariate model shows limited evidence of the covariates' relevance to population consumption, we also examined their relevance to per-adult consumption defined by (\ref{10th}). To this end, we considered the 5-covariate model and the no-covariate model. Results are shown in Fig.~\ref{fig-sr-adult} and Table \ref{table-adult}. Again, we observe that removing covariates from the model increases the credibility of $\rho_{sr}$, although at the expense of a minor loss of predictive performance in this case (DIC increased from 28.9 to 30.1). Comparing Figs.~\ref{fig-sr-adult} and \ref{fig-sr.links}, we see that trophic clustering according to feeding activity depends on whether feeding consumption is considered on a (i) per-adult basis or (ii) per-population basis. Specifically, relative positions of nodes in Graph SR are very different between (i) and (ii). Fig.~\ref{fig-sr-adult}, top left panel shows somewhat large uncertainty in activity level for $i$=1 (phytoplankton), 4 (benthic carnivores), 16 (horse mackerel), and 23 (kob) even though their posterior means are roughly farthest apart in the web. In contrast, Fig.~\ref{fig-sr.links}, top left panel shows that (ii) exhibits a clearer distinction in activity level among various nodes.%
}%

{%
However, unlike (ii), Table \ref{table-adult} shows that the tendency is moderately credible ($\ge$75\%) for (in)active feeders by volume to be (in)actively consumed by volume, and all covariates except taxonomic distance are at least moderately relevant to per-adult consumption ($\ge$70\% credible levels for $\beta$-intervals to exclude 0). This contrasts substantially with the weak relevance of covariates to population consumption. Also different between the two definitions of consumption is the sign of $\beta_5$, which is negative for per-adult consumption. Although still a non-causal relationship, the evidence for $\beta_5$$<$0 suggests that trophic species which are more heavily harvested tend to consume less on a per-adult basis and {\em vice versa}. Irrespective of the sign of $\beta_5$, the non-temporal nature of the data makes it impossible to use the current statistical SNA to address the potential effects of a seal cull (heavy harvest of seals) on commercial fisheries.%
}
\par\noindent
\begin{table}[!h]
\begin{center}
\caption{Benguela HPD intervals of interest for the reduced model as for Table \ref{table-full-intervals}, except for taking $y_{ij}$ as per-adult consumption (\ref{10th}). \label{table-adult}}
\begin{tabular}{lrrlc}\\\hline
&& \multicolumn{2}{c}{HPD interval} & \\ \cline{3-4}
Parameter & Posterior mean & Lower limit & Upper limit & Credible level \\ \hline
$\beta_1$ (taxo.~dist.) & 0.003 & 0.000 & \ \ \ 0.006 & 0.50 \\
$\beta_2$ (prey biomass) & $-$0.007 & $-$0.014 & $-$0.000 & {\bf 0.85} \\
$\beta_3$ (prey~harvest) & $-$0.004 & $-$0.009 & $-$0.000 & {\bf 0.70} \\
$\beta_4$ (pred.~biomass) & 0.013 & 0.001 & \ \ \ 0.025 & {\bf 0.99} \\
$\beta_5$ (pred.~harvest) & $-$0.005 & $-$0.009 & $-$0.000 & {\bf 0.70} \\
$\rho_{sr}$ & 0.402 & 0.077 & \ \ \ 0.993 & {\bf 0.75} \\
$\rho_{sr}$$^\text{a}$ & 0.483 & 0.124 & \ \ \ 0.988 & {\bf 0.80} \\ 
$\rho$ & $-$0.128 & \multicolumn{2}{c}{(interval includes 0)} & 0.50 \\ 
$\rho$$^\text{a}$ & $-$0.080 & \multicolumn{2}{c}{(interval includes 0)} & 0.50 \\ \hline
\end{tabular}\\
\footnotesize
\mbox{$^\text{a}$ From the no-covariate model.\hspace{3.8in}}
\end{center}
\end{table}
\section{Discussion}\label{sec-discuss}
{%
Throughout its long history, the analysis of trophic food web data has revolved around descriptive statistics such as connectance (a link density measure), mean food chain length, etc. Due to the static nature of studies from which most trophic data arise, quantitatively rigorous food web inference based on these statistics had been limited, even when various assumptions of feeding behavior are imposed through the classic cascade model \citep{cohen} or niche model \citep{williams}. \citet{winemiller}, while referring to the St.~Martin study in \citep{goldwasser} and food web studies in general, points out that ``[p]redator-prey interactions are highly variable, both in time and space, and this variation must be estimated and incorporated into descriptive and comparative food web research. [Yet, t]he empirical food web literature has been surprisingly uncritical toward estimates for trophic links. \ldots Precise, accurate, and reliable estimates of empirical food webs would open up exciting new avenues of comparative research. [Moreover, o]nly two of [some major] factors (predation/parasitism and food availability) [known to influence local population densities] are explicitly represented in food webs. [However, t]he food web structure \ldots of some, perhaps even most, ecosystems would be impossible to interpret or predict without fundamental understanding of abiotic drivers[.]''%
}

{%
Food webs are networks, and many existing SNA modeling approaches may be considered for adaptation to handling binary or weighted food web data. In this paper, we have proposed a statistical modeling framework for weighted food web data to address some of the above classic concerns in food web research. Built upon the SNA methodology by \citet{pnas} for 0-1 trophic data and associated covariates, our framework allows rigorous examination of inherently random weighted trophic relations from the perspectives of (i) feeding activity, (ii) preference of being prey, and (iii) preference for prey, all with respect to consumption on a per- individual/population basis while accounting for any observed biotic and abiotic driver(s) as covariate(s). This complements the framework in \citep{pnas} through which (i) to (iii) were considered with respect to the action of consuming and being consumed but irrespective of volume. While precision of the inference heavily depends on the quality of the data, our methodology does allow one to rigorously assess this precision through posterior distributions.%
}

{%
Our approach was used to study the consumption volumes in the Benguela system, and consumption frequencies in the St.~Martin systems. In both cases, we observed that the inference for two-dimensional sending and receiving preference spaces can be weak with a reduced web that is only a small subset of the complete web. Specifically, MCMC mixing is difficult, and the uncertainty for $\bs{u}_i$s and $\bs{v}_j$s can be overwhelmingly large for most nodes. Despite this, it appears that basal nodes and top predators can be clearly distinguishable from middle nodes with respect to feeding preference modeled as two-dimensional. In contrast, by reducing the dimensionality of the sending and receiving preference spaces to 1, MCMC mixing required little effort, although possibly at the expense of lowered model performance (according to the deviance information criterion) and reduced distinguishability among basal nodes in the sending preference space.%
}

{%
There are limitations to our framework for modeling weighted food web data. Indeed, the removal of $y_{ij}=$ 0 from consideration may not be ideal, as it drastically reduces the amount of available data. Consequently, inference for feeding preference can be weakened; the larger the web, the likelier this is the case. Furthermore, it is not straightforward to interpret phenomena such as $\rho_{sr}>$ 0 and $\rho>$ 0 in this unconventional context of trophic links. As Dr.~Beth Fulton of CSIRO Marine and Atmospheric Research (CMAR) pointed out at the 2010 CMAR Trophodynamics Workshop in Hobart, Tasmania, the existence of true zeros for links between pairs is often one of the most important aspects of food web research. Before a more innovative statistical framework is available to model weighted food web data that are heavily dominated by 0s, we have presented our statistical SNA technique for reduced webs as a compromise between relying solely on presence-absence data, and on modeling weighted $y_{ij}\geq$ 0 with complex distributional assumptions for $y_{ij}$.%
}

{%
Currently, we are pursuing an extension through a mixture model to incorporate the point mass of $y_{ij}$ at 0, whereby $p_{ij}$=$P(y_{ij}$$>$0$), p_{ij}$ follows (\ref{logistic}), $y_{ij}^\ast\,|\,\{y_{ij}>0\},\bs{x}_{ij},\bs{\beta}^\ast,\sigma^\ast \ \sim\text{N}(\bs{x}_{ij}'\bs{\beta}^\ast,{\sigma^\ast}^2)$, and $y_{ij}^\ast$ is the appropriately transformed value of $y_{ij}$ to achieve normality, and $\bs{\beta}^\ast$ and $\sigma^\ast$ are parameters at the level of $y_{ij}^\ast$ (not the level of $p_{ij}$). Variants of this model, including one with random slope parameters per node, are also being considered. When properly formulated and implemented, the SNA mixture model is expected to mitigate several existing difficulties associated with the reduced framework of this paper. However, new computational challenges may arise due to increased model complexity.
Finally, we also intend to address research questions that pertain to population dynamics, such as that on seal cull. For this, we expect to adapt the LSN framework of \citet{westveld} to the food web context for modeling any available proxy for temporal diet data.%
}

\section*{Acknowledgments}
{%
This paper was a result of research primarily conducted during AHW's Visiting Scientist position with CSIRO Mathematics, Informatics and Statistics (CMIS), for a project under the CMIS Capability Development Fund Theme. AHW thanks CMIS for sponsoring his visits. The authors thank Project Leader, Dr.~K.R.~Hayes, for his comments on a preliminary version of this paper; and various reviewers of a recent version for their valuable suggestions.%
}

\bibliographystyle{model3-num-names}

\section*{Figures}
\clearpage
\clearpage
\clearpage
\clearpage
\clearpage


\clearpage
\appendix
\section{Inference for preference space}\label{sec-proc}
{%
Although their cross-products are identifiable, individually the random effects corresponding to sender and receive preferences are unidentifiable without constraints. In the case of $k$=2, a two-set Procrustes transformation is employed in \citep{pnas} to circumvent this unidentifiability. This corresponds to obtaining, for each $t$th MCMC scan, a 2$\times$2 orthogonal transformation $\mathbb{Q}^{(t)}$ such that (a) $[\mathbb{Q}^{(t)}\bs{u}_i^{(t)}]'[\mathbb{Q}^{(t)}\bs{v}_j^{(t)}]=\bs{u}_i^{(t)\prime}\bs{v}_j^{(t)}$ for all $i\ne j$, and (b) $\mathbb{Q}^{(t)}$ minimizes $||\mathbb{U}_0-\mathbb{Q}^{(t)}\mathbb{U}^{(t)}||+||\mathbb{V}_0-\mathbb{Q}^{(t)}\mathbb{V}^{(t)}||$, the total Euclidean distance from the $t$th orientation to the default orientation of both preference spaces. Specifically, the $t$th orientation of sender preference is spanned by the columns of $\mathbb{U}^{(t)}=[\bs{u}_1^{(t)}\ldots\bs{u}_n^{(t)}]$, and the corresponding default orientation is spanned by the columns of $\mathbb{U}_0=[\bs{u}_{1,0}\ldots\bs{u}_{n,0}]$, where $\mathbb{U}_0$ is arbitrarily defined at $\bs{u}_{i0}=(1/T)\sum_{t=1}^T\bs{u}_i^{(t)}$; and similarly for receiver preference spanned by the columns of $\mathbb{V}^{(t)}$ or $\mathbb{V}_0$. The approach stems from least-squares optimization, and does not directly reduce to the case of $k$=1 due to degeneracy when $\mathbb{Q}$ is scalar.%
}

{%
Instead, for $k$=1, one can arbitrarily fix $u_i$ or $v_i$ for some $i\in{\cal I}_2$ at a non-zero value. In the Benguela web, for example, $i$=3 (bacteria) is a middle node, so that we can incorporate the constraint $u_3\equiv$ 1 into the model. As $u_i v_j$ is identifiable for all $\{(i,j):i\neq j\text{ and }y_{ij}>0\}$, this constraint imposes identifiability on $v_j$ for all $\{j:j\ne 3\text{ and }y_{3j}>0\}$. Identifiability of $v_{j'}$ for each $j'$ from this set in turn imposes identifiability on $u_i$ for all $\{i:i\ne 3,j'\text{ and }y_{ij'}>0\}$, etc.%
}

{%
All modeling results for $k$=1 in this article correspond to $u_3\equiv$ 1 for Benguela and $u_5\equiv$ 1 for St.~Martin (Node 5 is {\it Elaenia}, a type of flycatcher).%
} 

\section{Prior distributions and MCMC mixing concerns}\label{sec-prior.mix}
For Bayesian inference of complex models, the mixing of MCMC draws is often of practical concern. To mitigate mixing difficulties, first note that (\ref{sr-mutual}) implies
\begin{equation} \label{sr-transf}
s_i\sim\text{N}(0,\sigma_s^2)\,,\hspace{1cm}
r_i| s_i\sim\text{N}(\lambda s_i,\phi^2)\,,\hspace{1cm}
\sigma_r^2=\phi^2+\lambda^2\sigma_s^2\,, \hspace{1cm}
\rho_{sr}=\lambda\frac{\sigma_s}{\sigma_r}\,. \hspace{1cm}
\end{equation}
To see this, rewrite the last expression of (\ref{sr-transf}) as $\lambda = \rho_{sr}\sigma_r/\sigma_s$. Hence, the second distribution in (\ref{sr-transf}) implies $\phi^2=(1-\rho_{sr}^2)\sigma_r^2$. These equations for $\lambda$ and $\phi^2$ imply $\rho_{sr}^2 = \lambda^2\sigma_s^2/\sigma_r^2 = (\sigma_r^2-\phi^2)/\sigma_r^2$, which in turn implies the expression for $\sigma_r^2$ in (\ref{sr-transf}). The use of (\ref{sr-transf}) avoids generating MCMC samples of $\bs{\Sigma}$ from a matrix distribution such as Wishart, which caused major mixing difficulties in our case. Mixing issues associated with Wishart priors are also discussed in \cite{bugs}.

We consider reasonably diffuse proper priors in the Bayesian hierarchy. The diffuseness reflects our lack of prior knowledge of parameter values. Let $\Gamma(a,b)$ denote the Gamma distribution parametrized in such a way that small values of $a$ and $b$ lead to diffuseness. We take
\par\noindent%
\begin{eqnarray}
\rho=\frac{e^{2z}-1}{e^{2z}+1}\,,\hspace{1cm}
z\sim\text{N}(0,0.82^2)\,, \label{fisher}\\
\lambda,\beta_\ell\sim\text{N}(0,a^{-1})
\text{\hspace{1cm}for all }\ell=0,1,\ldots,L\,, \nonumber \\
\sigma^{-2},\sigma_s^{-2},\phi^{-2},\sigma_{uq}^{-2},\sigma_{vq}^{-2}\sim\Gamma(a,a)
\text{\hspace{1cm}for all }q=1,\ldots,k\,, \nonumber
\end{eqnarray}
where $L$ is the number of covariates ($L$ is up to 5 for Benguela and up to 1 for St.~Martin), $a\in\{0.1,0.01,0.001\}$, and $k=$ 1 or 2. The actual choice of $a$ varied in our analyses and was found to have virtually no influence on the results. Expression (\ref{fisher}) employs the Fisher transformation to avoid taking $\rho\sim\text{U}[-1,1]$ so as to improve MCMC mixing; our choice of distribution for $z$ corresponds very closely to $\rho\sim\text{U}[-1,1]$. Note that $\rho$ is not modeled for St.~Martin because ${\cal S}_1=\emptyset$.

\section{Supplementary tables and figures}
\setcounter{figure}{0}
\begin{figure}[!h]
\begin{center}
\caption{Scatter plot of positive population consumption (\ref{20th}) versus transformed predator harvest $t_{j2}^{r\ast}$. Vertical line corresponds to seals as predators. Highlights correspond to three species that are prey to African fur seals' and are of commercial interest: red ``a'' denotes anchovy, green ``m'' denotes horse mackerel, and blue ``h'' denotes hakes. \label{fig-y-vs-predharv}}
\makebox{\includegraphics[scale=.6]{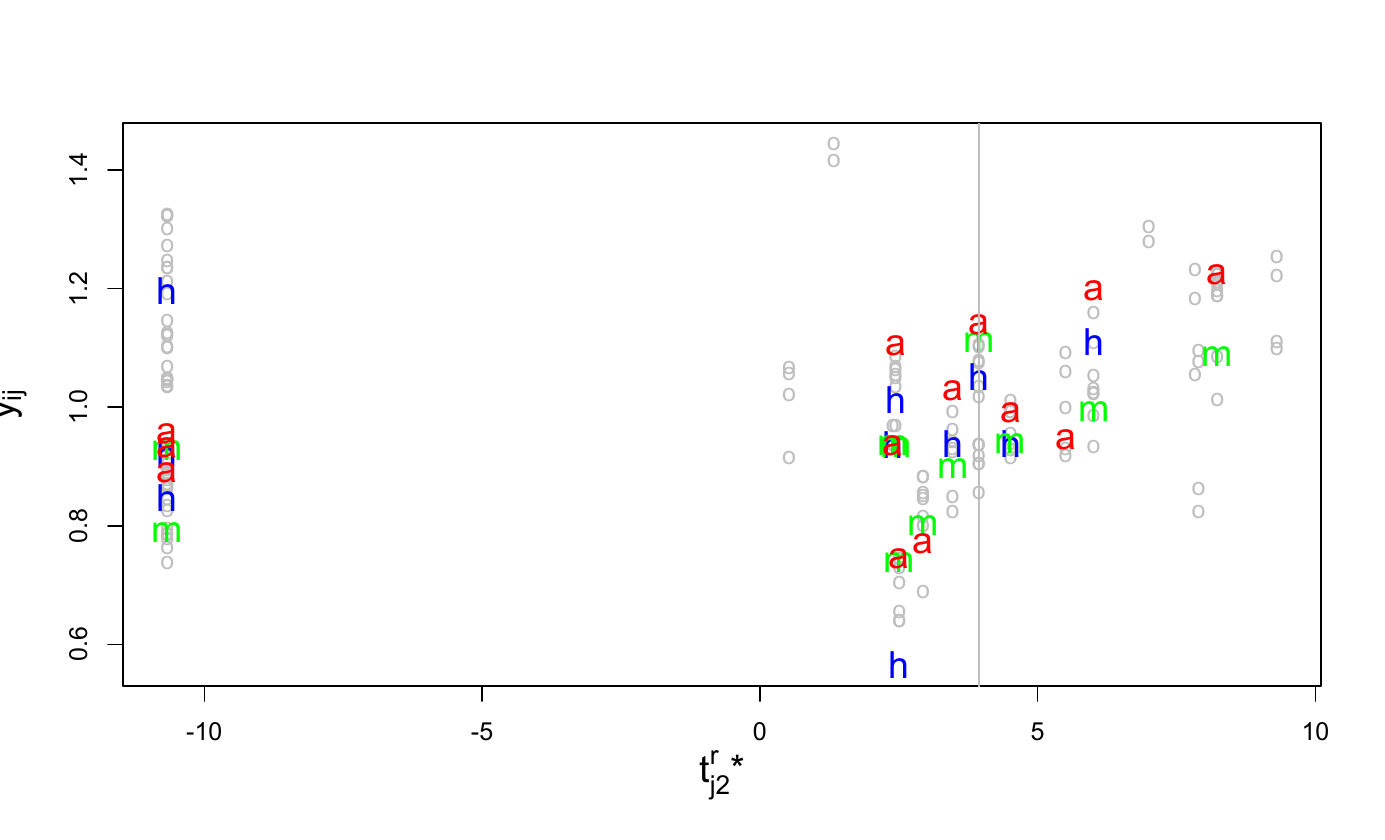}}
\end{center}
\end{figure}
\setcounter{table}{0}
\begin{table}[!h]
\begin{center}
\caption{The Benguela web. Source: \citet{yodzis}.\label{nodes}}
\begin{tabular}{rlrl}\hline
Node & Organism(s) & Node & Organism(s) \\ \hline
1 & Phytoplankton$^\text{a}$ &
2 & Benthic filter-feeders$^\text{a}$ \\
3 & Bacteria &
4 & Benthic carnivores \\
5 & Microzooplankton &
6 & Mesozooplankton \\
7 & Macrozooplankton &
8 & Gelatinous zooplankton \\
9 & Anchovy &
10 & Pilchard \\
11 & Round herring &
12 & Lightfish \\
13 & Lanternfish &
14 & Goby \\
15 & Other pelagics &
16 & Horse mackerel \\
17 & Chub mackerel &
18 & Other groundfish \\
19 & Hakes &
20 & Squid \\
21 & Tunas &
22 & Snoek \\
23 & Kob &
24 & Yellowtail \\
25 & Geelbek &
26 & Whales and dolphins \\
27 & Birds &
28 & Seals \\
29 & Sharks \\ \hline
\end{tabular}\\
\footnotesize
\mbox{$^\text{a}$ Basal.\hspace{3.5in}}
\end{center}
\end{table}
\begin{table}[!h]
\begin{center}
\caption{The St.~Martin web. Source: \citet{goldwasser}.\label{nodes2}}
\begin{tabular}{rlrl}\hline
Node & Organism(s) & Node & Organism(s) \\ \hline
1 & Kestrel$^\text{a}$ &
2 & Scaly-breasted thrasher$^\text{a}$ \\
3 & Pearly-eyed thrasher$^\text{a}$ & 
4 & Gray kingbird$^\text{a}$ \\
5 & {\it Elaenia} & 
6 & Yellow warbler$^\text{a}$ \\
7 & Bullfinch & 
8 & Grassquit \\
9 & Hummingbirds &
10 & Bananaquit \\
11 & {\it Thelandros cubensis}$^\text{a, b}$ & 
12 & {\it Mesocoelium} sp.$^\text{a, b}$ \\ 
13 & {\it Alogyptus crenshawi}$^\text{a, b}$ & 
14 & {\it Anolis gingivinus} \\
15 & {\it Anolis pogus} & 
16 & Collembola \\
17 & Orthoptera & 
18 & Isoptera \\
19 & Hemiptera &
20 & Homoptera \\
21 & Thysanoptera & 
22 & Coleoptera adult \\
23 & Coleoptera larva & 
24 & Ants \\
25 & Other hymenoptera & 
26 & Lepidoptera adult \\
27 & Lepidoptera larva & 
28 & Diptera adult \\
29 & Diptera larva & 
30 & Adult spider \\
31 & Juvenile spider & 
32 & Annelid \\
33 & Gastropoda & 
34 & Mites \\
35 & Centipede & 
36 & Millipede \\
37 & Isopoda & 
38 & Fungi \\
39 & Fruit and seeds$^\text{c}$ &
40 & Nectar and floral$^\text{c}$ \\
41 & Leaves$^\text{c}$ & 
42 & Wood$^\text{c}$ \\
43 & Roots$^\text{c}$ & 
44 & Detritus$^\text{c}$ \\ \hline
\end{tabular}\\
\footnotesize
\mbox{$^\text{a}$ Top predator.\hspace{3.6in}} \\
\mbox{$^\text{b}$ Parasitic.\hspace{3.8in}} \\
\mbox{$^\text{c}$ Basal.\hspace{3.95in}}
\end{center}
\end{table}
\begin{table}[!h]
\begin{center}
\caption{Benguela model comparison with predator harvest as sole covariate.\label{dic}}
\begin{tabular}{ccr}\\\hline
Slope & Include $\bs{u}_i'\bs{v}_j$? & DIC \\ \hline
Fixed & Yes & $-$245 \\
Fixed & No & $-$217 \\
Random & No & $-$211 \\
Random & Yes & $-$188 \\ \hline
\end{tabular}
\end{center}
\end{table}
\begin{table}[!h]
\begin{center}
\caption{HPD intervals for slopes $\beta_{i5}$ in the Benguela random-slope model without $\bs{u}_i'\bs{v}_j$ (Section \ref{sec-cull}). Credible levels above 0.5 are presented in bold. Omitted are $i$'s for which corresponding 50\% HPD intervals include 0.\label{random-slope}}
\begin{tabular}{rrrrc}\\\hline
&& \multicolumn{2}{c}{HPD interval} & \\ \cline{3-4}
$i$ & Posterior mean & Lower limit & Upper limit & Credible level \\ \hline
1 & $-$0.035 & $-$0.058 & $-$0.010 & {\bf 0.99} \\
4 & $-$0.007 & $-$0.013 & $-$0.000 & {\bf 0.75} \\
8 & 0.009 & 0.002 & 0.017 & {\bf 0.75} \\
9 & $-$0.006 & $-$0.013 & $-$0.000 & {\bf 0.70} \\
10 & 0.008 & 0.000 & 0.015 & {\bf 0.85} \\
12 & 0.007 & 0.001 & 0.013 & {\bf 0.65} \\
14 & $-$0.005 & $-$0.010 & $-$0.000 & {\bf 0.60} \\
15 & 0.016 & 0.002 & 0.030 & {\bf 0.85} \\
17 & 0.009 & 0.000 & 0.017 & {\bf 0.75} \\
18 & 0.008 & 0.001 & 0.016 & {\bf 0.95} \\
19 & 0.006 & 0.000 & 0.012 & {\bf 0.75} \\
20 & $-$0.003 & $-$0.008 & $-$0.000 & {\bf 0.60} \\
22 & 0.012 & 0.001 & 0.023 & {\bf 0.90} \\
23 & 0.006 & 0.000 & 0.013 & {\bf 0.65} \\
24 & 0.009 & 0.000 & 0.017 & {\bf 0.75} \\
29 & 0.008 & 0.001 & 0.016 & 0.50 \\ \hline
\end{tabular}
\end{center}
\end{table}

\end{document}